\documentclass[aps,showpacs,floatfix,twocolumn,superscriptaddress,longbibliography]{revtex4-2}

\usepackage{amsmath}
\usepackage{amssymb}
\usepackage{amstext}
\usepackage{amsopn}
\usepackage{amsfonts}
\usepackage{amsxtra}
\usepackage[english]{babel}
\usepackage{graphicx}
\usepackage{float}
\usepackage{bm}
\usepackage{multirow}
\usepackage{mathrsfs}
\usepackage{mathtools}
\usepackage{dcolumn}
\usepackage[colorlinks]{hyperref}
\usepackage{color}
\usepackage{url}
\usepackage{breakurl}
\usepackage{relsize}

\begin{document}
\title{Optical properties and carrier localization in the layered phosphide EuCd$_\mathbf{2}$P$_\mathbf{2}$}
\preprint{Draft, not for distribution}
\author{C. C. Homes}
\email[]{homes@bnl.gov}
\affiliation{National Synchrotron Light Source II, Brookhaven National Laboratory, Upton, New York 11973, USA}
\author{Z.-C. Wang}
\author{K. Fruhling}
\author{F. Tafti}
\email[]{fazel.tafti@bc.edu}
\affiliation{Department of Physics, Boston College, Chestnut Hill, Massachusetts 02467, USA}
\date{\today, version 2.5}

%
%
\begin{abstract}
The temperature dependence of the complex optical properties of the layered phosphide material
EuCd$_2$P$_2$ have been measured over a wide frequency range above and below $T_{\rm N} \simeq
11.5$~K for light polarized in the \emph{a-b} planes.  At room temperature, the optical
conductivity is well described by a weak free-carrier component with a Drude plasma frequency
of $\simeq 1100$~cm$^{-1}$ and a scattering rate of $1/\tau_D\simeq 700$~cm$^{-1}$, with the
onset of interband absorptions above $\simeq 2000$~cm$^{-1}$.  Two infrared-active $E_u$ modes
are observed at $\simeq 89$ and 239~cm$^{-1}$.
As the temperature is reduced the scattering rate decreases and the low-frequency conductivity
increases slightly; however, below $\simeq 50$~K the conductivity decreases until at the
resistivity maximum at $\simeq 18$~K (just below $2T_{\rm N}$) the spectral weight
associated with free carriers is transferred to a localized excitation at $\simeq 500$~cm$^{-1}$.
Below $T_{\rm N}$, metallic behavior is recovered.  Interestingly, the $E_u$ modes are largely
unaffected by these changes, with only the position of the high-frequency mode showing any
signs of anomalous behavior.
While several scenarios are considered, the prevailing view is that the resistivity maximum
and subsequent carrier localization is due to the formation of ferromagnetic domains below
$\simeq 2T_{\rm N}$  that result in spin-polarized clusters due to spin-carrier coupling
\cite{Sunko2022}.
\end{abstract}
%
%
%
\maketitle

%
%
%
\section{Introduction}
Magnetic semimetals display a variety of interesting phenomena.  The layered europium
materials EuCd$_{2}X_2$, with $X=$Sb, As, and P, are of particular interest as the Sb and As
materials are magnetic Weyl semimetals \cite{Wang2016,Ma2019,Wang2019,Behrends2019,
Soh2019,Jo2020,Su2020,Ma2020,Xu2021}.  These hexagonal materials have a layered crystal structure,
with the Cd$_{2}X_2$ layers separated by the europium layers; the magnetism originates from
the europium layers which order antiferromagnetically at low temperature
\cite{Schellenberg2011,Rahn2018}.
The transport properties of these materials are intriguing as they display peaks in the
resistivity close to the N\'{e}el temperature.  In the case of As ($T_{\rm N}\simeq 9.5$~K),
the in-plane resistivity roughly triples, while in the case of Sb ($T_{\rm N}\simeq 7.4$~K),
it is more of a shoulder-like feature; however, in both cases the increase occurs at
$T_{\rm N}$ and  is suppressed by the application of a modest magnetic field \cite{Soh2020}.
This behavior is dramatically exaggerated in the phosphide material ($T_{\rm N}\simeq 11.5$~K).
At room temperature, the in-plane resistivity may be described as that of a poor metal with
$\rho_{ab}\simeq25$~m$\Omega\,{\rm cm}$; this value increases by roughly two orders
of magnitude well above $T_{\rm N}$ at $\simeq 18$~K, effectively rendering the sample
semiconducting; below this temperature the resistivity drops dramatically, falling slightly
below the room temperature value at and below $T_{\rm N}$ \cite{Wang2021}.  As with Sb and
As, the resistivity peak is suppressed with magnetic field, resulting in a colossal
magnetoresistance.  Key questions for the phosphide material are: What is the nature of
the free-carrier response, and what becomes of the free-carriers at the resistivity
maximum?
The optical conductivity is ideally suited to address these issues. The change in the
dc conductivity from $\sigma_{dc}\simeq 65\,\Omega^{-1}{\rm cm}^{-1}$ at $\simeq 50$~K to a
value that is effectively zero from an optical point of view should have a dramatic
signature in the optical properties.  The frequency dependence of the optical conductivity
will also allow the spectral weight and the scattering rate associated with the free carriers
to be determined, as well as the shifts in the spectral weight near the resistivity maximum
to be tracked.

%
%
In this work the temperature dependence of the in-plane complex optical properties have
been determined for a single crystal of EuCd$_2$P$_2$ over a wide frequency range for
light polarized in the \emph{a-b} planes.  At room temperature, two infrared-active
vibrational modes and a weak Drude-like free-carrier component are superimposed on an
otherwise semiconducting response with an onset of absorption due to interband transitions
at $\gtrsim 2000$~cm$^{-1}$.
While the spectral weight associated with the free carriers remains roughly constant
with decreasing temperature, the scattering rate decreases by about 10\% between 295
and 50~K, consistent with the transport values for the resistivity.  Below 50~K,
the resistivity (and the scattering rate) increases by roughly a factor of two at 25~K,
heralding the complete disappearance of the free-carrier component, the totality of
which is transferred into a localized excitation centered at $\simeq 500$~cm$^{-1}$
($\simeq 62$~meV); below $T_{\rm N}$ the metallic behavior is restored and the spectral
weight associated with the localized excitation is transferred back into the free-carrier
response.
The dramatic changes in the resistivity and the optical properties are intimately
connected to the magnetism in this material.  While several different possibilities are
considered, it appears that formation of ferromagnetic clusters at $\simeq 2\,T_{\rm N}$
result in carrier localization in spin-polarized clusters \cite{Sunko2022}; this effect
is suppressed when the clusters become contiguous and form a network.  Moreover, below
$T_N$ the conductivity increases slightly, likely due to a decrease in fluctuations.

%
%
\begin{figure}[tb]
\centerline{\includegraphics[width=2.95in]{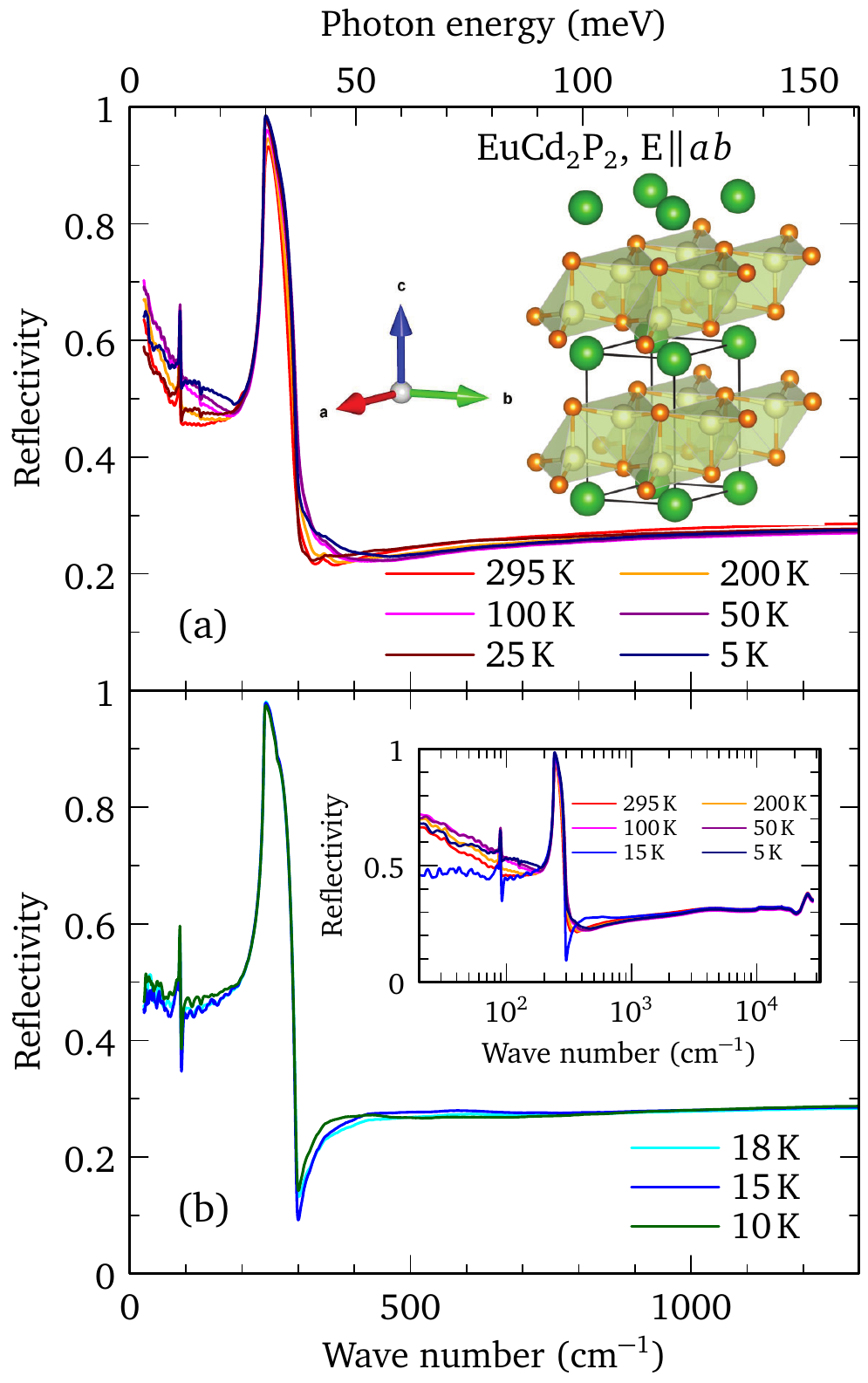}}%
\caption{(a) The temperature dependence of the reflectivity of a single crystal of EuCd$_2$P$_2$
versus wave number in the far-infrared region for light polarized in the \emph{a-b} planes for
temperatures between 25 and 295~K, as well as at 5~K, where the response of the reflectivity
is metallic.  The resolution at low frequency is typically better than 2~cm$^{-1}$.
Inset: the unit cell of EuCd$_2$P$_2$ showing the cadmium-phosphide layers, separated by
the Eu layers \cite{VESTA}.
(b) The reflectivity between 10 and 18~K for light polarized parallel to the \emph{a-b} planes,
where the response is now indicative of an insulating or semiconducting behavior.
Inset: the reflectivity on a semi-log plot at several temperatures shown over a much wider frequency range.
}
\label{fig:reflec}
\end{figure}

%
%
\section{Experiment and Results}
Single crystals of EuCd$_2$P$_2$ were grown using a flux technique that has been previously
described \cite{Wang2021}; x-ray diffraction on the large, mirror-like crystal faces
revealed that they contain the \emph{a-b} planes.  It should be noted that there is some
terracing on the surface of the crystal.
The reflectivity of an as-grown crystal face ($\simeq 1.5\,{\rm mm}\times 2\,{\rm mm}$) has been
measured at a near-normal angle of incidence for light polarized parallel to the \emph{a-b}
planes over a wide temperature and frequency range ($\simeq 2$~meV to 4~eV) using an
\emph{in situ} evaporation technique \cite{Homes1993}; the results are shown in Fig.~\ref{fig:reflec}.

The character of the reflectivity may be described as either poorly metallic, or semiconducting
(insulating), depending on the temperature.  The response of the reflectivity is shown in the
metallic case in Fig.~\ref{fig:reflec}(a) at temperatures between 25 and 295~K, and at 5~K;
for these temperatures the low-frequency reflectivity is increasing rapidly with decreasing frequency,
which is in agreement with the requirement that when $\sigma_{dc}\neq 0$, $R(\omega \rightarrow 0) = 1$.
The two prominent features in the reflectivity that are attributed to the normally infrared-active
lattice modes are also partially screened \cite{Homes2000}.
In contrast, the reflectivity in Fig.~\ref{fig:reflec}(b) between 10 and 18~K shows a dramatically
different response; $R(\omega\rightarrow 0) \simeq {\rm const}$, the result expected for an insulator
or semiconductor.  In addition, the two lattice modes appear to be almost totally unscreened,
resulting in fundamentally different line shapes \cite{SupMat}.

%
%
The reflectivity is a combination of the real and imaginary parts of the dielectric function,
and as such can be difficult to interpret; the real part of the optical conductivity, calculated
from the imaginary part of the dielectric function, is a more intuitive quantity.
Accordingly, the complex dielectric function, $\tilde\epsilon(\omega) =
\epsilon_1 + i\epsilon_2$, has been determined from a Kramers-Kronig analysis of the
reflectivity \cite{Dressel-Book}, which requires extrapolations at high and low frequency.
In the case where metallic conductivity is observed, at low frequency a metallic Hagen-Rubens
extrapolation, $R(\omega) = 1 - A\sqrt{\omega}$ was employed, where $A$ is chosen to match
the value of the reflectance at the lowest measured frequency.  Where an insulating or
semiconducting response is observed, below the lowest measured frequency the reflectance
was assumed to be constant.  Above the highest-measured frequency point the reflectance
was assumed to follow a $\omega^{-1}$ dependence up to $8\times 10^4$~cm$^{-1}$, above which
a free-electron approximation ($R\propto \omega^{-4}$) was assumed \cite{Wooten}.
The complex conductivity, $\tilde\sigma(\omega)$, is calculated from from the complex
dielectric function, $\tilde\sigma(\omega) = \sigma_1 +i\sigma_2 = -2\pi i \omega
[\tilde\epsilon(\omega) - \epsilon_\infty ]/Z_0$, where $\epsilon_\infty$ high-frequency
contribution to the real part of the dielectric function, and $Z_0\simeq 377$~$\Omega$ is
the impedance of free space.

%
%
\begin{figure}[tb]
\centerline{\includegraphics[width=3.08in]{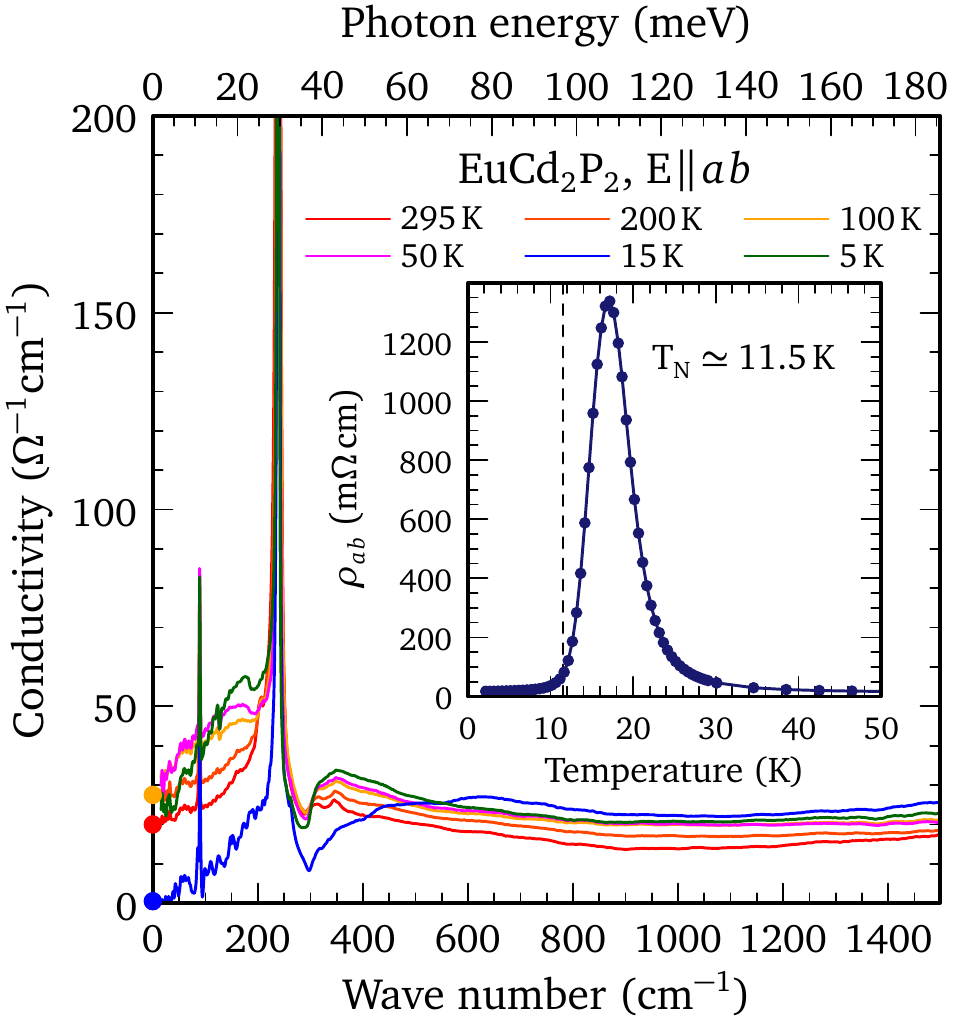}}%
\caption{The temperature dependence of the real part of the optical conductivity of EuCd$_2$P$_2$
versus wave number for light polarized in the \emph{a-b} planes in the low-frequency region.  This
material is a poor metal; however, at 15~K the response is semiconducting.  Two sharp lattice modes
are observed at $\simeq 89$ and 239~cm$^{-1}$, along with a notch-like feature at about 300~cm$^{-1}$.
The filled circles on the conductivity axis represent the extrapolated values for $\sigma_1(\omega
\rightarrow 0$).
Inset: the temperature dependence of the \emph{a-b} plane dc resistivity measured in zero magnetic field.
Note that the resistivity maximum occurs at $\simeq 18$~K, nearly twice the value of $T_{\rm N}\simeq 11$~K
\cite{Wang2021}.
}
\label{fig:sigma}
\end{figure}

The temperature dependence of the real part of the optical conductivity for EuCd$_2$P$_2$ in
the low-frequency region is shown in Fig.~\ref{fig:sigma} for light polarized in the \emph{a-b} planes.
At room temperature, the optical conductivity is that of a poor metal, with $\sigma_1(\omega\rightarrow 0)
\simeq 20\,\Omega^{-1} {\rm cm}^{-1}$, which is about a factor of two lower than the value obtained from
transport \cite{Wang2021}.  Two sharp features, one weak and the other strong, are observed at about
89 and 239~cm$^{-1}$, respectively.
As the temperature is reduced, the low-energy conductivity increases slightly, with
$\sigma_1(\omega\rightarrow 0) \simeq 30$~$\Omega^{-1}{\rm cm}^{-1}$ at 50 K, which is
consistent with the decreasing resistivity observed in transport \cite{Wang2021},
shown in the inset in Fig.~\ref{fig:sigma}.  At 10, 15 and 18~K (for clarity only the
conductivity at 15~K is shown in Fig.~\ref{fig:sigma}), there is a dramatic decrease in
the low-frequency conductivity, with a commensurate transfer of spectral weight
(area under the conductivity curve) to high frequency, leading to an increase in
the optical conductivity above $\simeq 600$~cm$^{-1}$.  Interestingly, below $T_{\rm N}$
at 5~K the metallic behavior is recovered and spectral weight is transferred back into the
free-carrier component.

%
%
\section{Discussion}
To investigate the behavior of the free-carriers in more detail, the optical response has been
modeled using the Drude-Lorentz model for the complex dielectric function
\begin{equation}
  \tilde\epsilon(\omega) = \epsilon_\infty - \frac{\omega_{p,D}^2}{\omega^2+i\omega/\tau_D}
    + \sum_j \frac{\Omega_j^2}{\omega_j^2 - \omega^2 - i\omega\gamma_j}.
  \label{eq:dl}
\end{equation}
In the first term $\omega_{p,D}^2 = 4\pi ne^2/m^\ast$ and $1/\tau_D$ are
the square of the plasma frequency and scattering rate for the delocalized (Drude)
carriers, respectively, and $n$ and $m^\ast$ are the carrier concentration and effective mass.
In the summation,  $\omega_j$, $\gamma_j$ and $\Omega_j$ are the position, width, and
strength of a symmetric Lorentzian oscillator that describe the $j$th vibration or bound
excitation.

\begin{figure}[t]
\centerline{\includegraphics[width=3.0in]{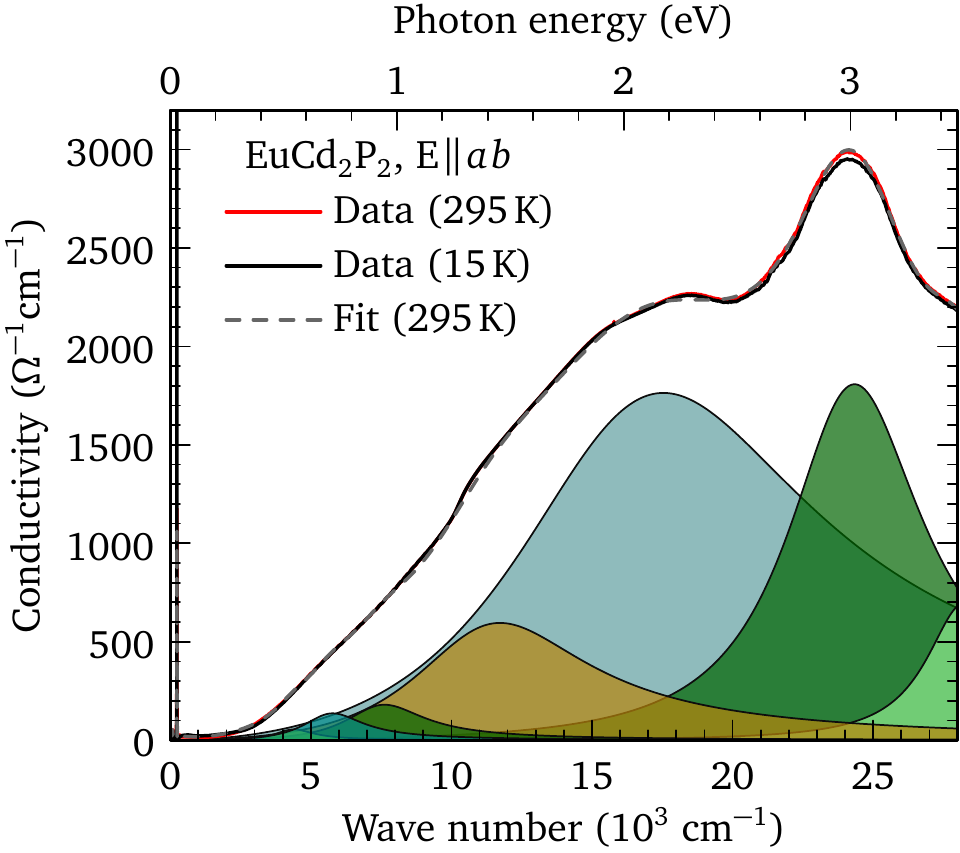}}%
\caption{The optical conductivity of EuCd$_2$P$_2$ at 295 and 15~K for light polarized
in the \emph{a-b} planes over a wide frequency range, showing the onset of absorption due to
interband transitions at $\simeq 2000$~cm$^{-1}$, and the lack of any strong temperature dependence
in this region.
The result of the fit to the optical conductivity of EuCd$_2$P$_2$ at 295~K using the
Drude-Lorentz mode is indicated by the dashed line; the fit is decomposed into the
contributions from the different Lorentz oscillators.}
\label{fig:inter}
\end{figure}

The strategy that we will adopt in fitting the optical conductivity is to first fit the
interband transitions to Lorentzian oscillators using a non-linear least squares technique,
then, either holding these high-frequency oscillators fixed or allowing only modest refinement,
fit the free-carrier response in the low-frequency region.  The general approach is to use
the minimum number of oscillators required to describe the data.
The real part of the in-plane optical conductivity is shown over a wide frequency range
at several temperatures in Fig.~\ref{fig:inter}, showing the onset of absorption above
$\simeq 2000$~cm$^{-1}$, along with the results of the fit at 295~K.  Oscillators have
been introduced at $\simeq 0.54$, 0.71, 0.95, 1.45, 2.17, 3.02, and 3.50~eV; the fit has
been decomposed into the contributions from the Lorentz oscillators. While the overall
fit is excellent, it can be argued that the low-frequency oscillator could probably be
removed and a reasonable fit still obtained; however, the low-frequency oscillator is
required to reproduce the linear region of the optical conductivity, which then allows
the low-frequency conductivity to be fit reliably.  We also note that because the oscillator
at 3.5~eV is at the limit of the measured data, it should be treated with caution.

%
%

The determination of the frequencies for the interband transitions allows the
low-frequency conductivity to be fit using the Drude model.  However, the sharp features
attributed to the lattice modes complicate this approach; it is simpler to fit these
features and then subtract them from the conductivity, leaving only the electronic
continuum associated with the free carriers.  Accordingly, the rest of the discussion
will first deal with the vibrational properties, followed by an analysis of the free-carrier
response.

%
%
\subsection{Vibrational properties}
\begin{figure}[t]
\centerline{\includegraphics[width=3.0in]{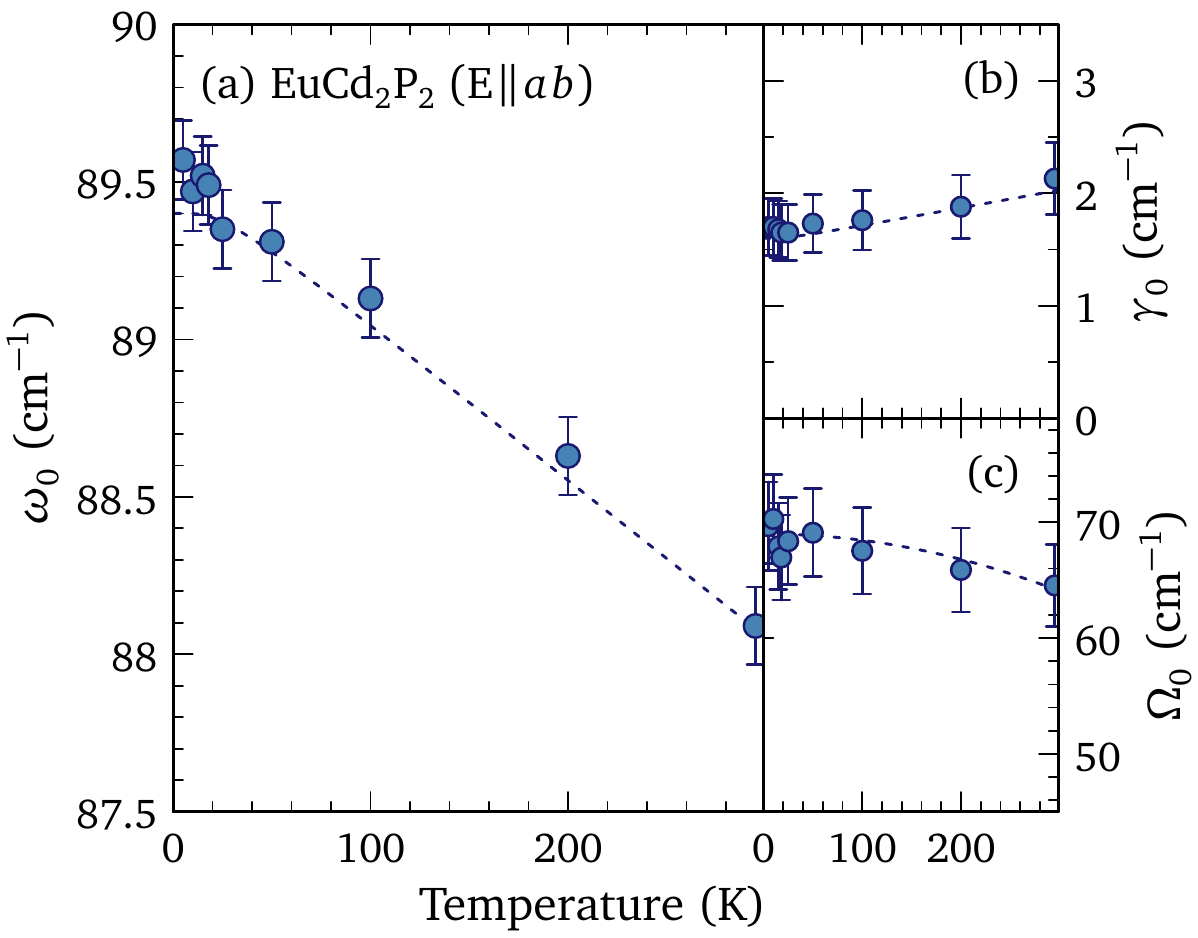}}
\vspace*{2.0mm}
\centerline{\includegraphics[width=3.1in]{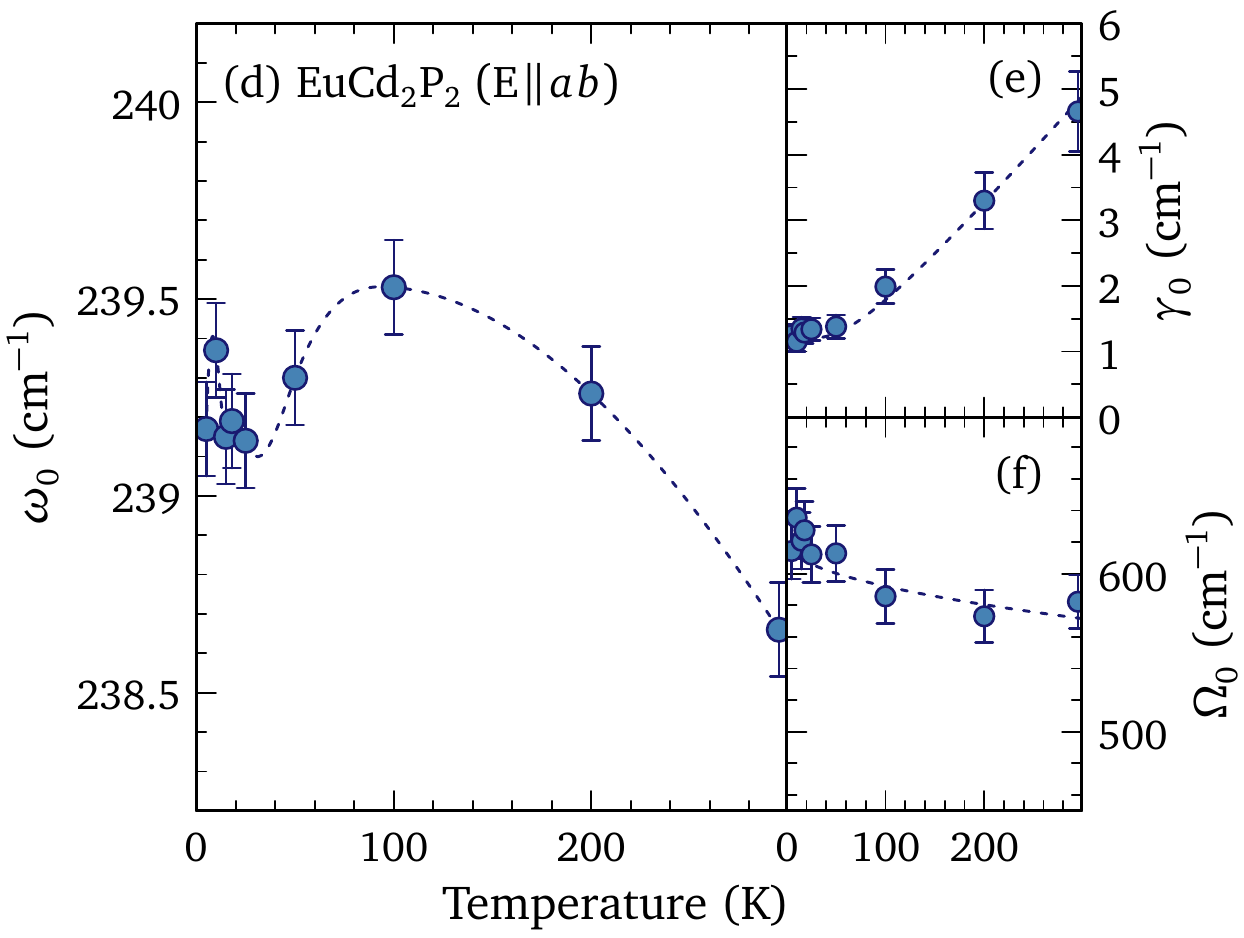}}
\caption{The results of the fit to the -n-plane  optical conductivity using
symmetric Lorentzian oscillators for the $E_u$ modes in EuCd$_2$P$_2$.
The upper panel shows the temperature dependence of (a) the position,
(b) line width, and (c) strength of the low-frequency $E_u$ mode; the
dashed lines for the position and line width are calculated using the
anharmonic decay model using $\omega_{\rm B}=89.5$~cm$^{-1}$ ($C=0.0019$),
and $\gamma_0=1.6$~cm$^{-1}$ ($\Gamma=0.032$), respectively.
The lower panel shows the (d) the position, (e) line width, and
(f) strength of the high-frequency $E_u$ mode.  The dashed line for the
line width is generated using the anharmonic decay model with $\gamma_0 =
1.2$~cm$^{-1}$ ($\Gamma=1.32$); all others are
drawn as a guide to the eye.}
\label{fig:modes}
\end{figure}

In the hexagonal (trigonal) $P\bar{3}m1$ setting, the irreducible vibrational representation
for EuCd$_2$P$_2$ is $\Gamma_{\rm irr} = 2A_{1g}+2E_g+2A_{2u}+2E_u$.  The $A_{1g}$ and
$E_g$ modes are Raman active, while the $A_{2u}$ and $E_u$ modes are infrared active
along the \emph{c} axis and the \emph{a-b} planes, respectively.  The two modes observed
in Fig.~\ref{fig:sigma} at $\simeq 89$ and 239~cm$^{-1}$ are the expected $E_u$ modes.
It is tempting to assume that the notch-like feature just above the high-frequency mode
at $\simeq 300$~cm$^{-1}$ is due to electron-phonon coupling resulting in a Fano-like
antiresonance in the electronic continuum \cite{Fano1961,Damascelli1996}.  However,
it should be noted that this vibration is exceptionally strong and narrow; it does not show
the broadening that would be expected for a mode that was coupled to the electronic
background \cite{Wang2017}.  Moreover, there is almost no electronic background for
it to interact with. The low-frequency mode also displays no sign of any asymmetry.
The two modes have therefore been fit using a symmetric Lorentzian on a linear background.
The results of the fits to the two $E_u$ mode are shown in Fig.\ref{fig:modes}.

%
%

The temperature dependence of the position of the low-energy $E_u$ mode, shown in
Fig.~\ref{fig:modes}(a), behaves in the expected way, increasing in frequency (hardening)
with decreasing temperature from $\simeq 88$ to about 89.5~cm$^{-1}$ at low temperature.
At room temperature this mode is quite narrow with a line width of $\simeq 2$~cm$^{-1}$,
and it decreases only slightly to 1.7~cm$^{-1}$ at low temperature; the oscillator strength
is roughly constant with $\Omega_0 \simeq 68$~cm$^{-1}$.  Overall, this behavior is what is
expected for a symmetric anharmonic decay of an optic mode into two acoustic modes with
identical frequencies and opposite momenta \cite{Klemens1966,Menendez1984}.  The functional
form employed here is,
\begin{equation}
  \omega_{0}(T)=\omega_{\rm B} \left[1-\frac{2{\rm C}}{e^x-1}  \right],
\end{equation}
\begin{equation}
  \gamma_{0}(T)=\Gamma_{0} \left[1+\frac{2\Gamma}{e^x-1} \right],
\end{equation}
where $\omega_{\rm B}$ is the bare phonon frequency, $\Gamma_{0}$ is a residual line
width, C and $\Gamma$ are constants, and $x=\hbar\omega_{\rm B}/(2k_{\rm B}T)$;
the bare phonon frequency (residual line width) is recovered in the $T\rightarrow 0$
limit \cite{Homes2016}.  The model fits are indicated by the dashed lines in
Figs.~\ref{fig:modes}(a) and \ref{fig:modes}(b).

In contrast, the frequency dependence of the high-frequency $E_u$ mode, shown in
Fig.~\ref{fig:modes}(d), is somewhat anomalous, initially hardening with decreasing
temperature, reaching a maximum of $\simeq 239.5$~cm$^{-1}$ at 100~K before softening
by nearly 1~cm$^{-1}$ as the temperature continues to decrease, then hardening again below
about 25~K, suggesting a weak coupling to the magnetism in this material.  Surprisingly,
this mode narrows from $\simeq 5$ to 1.3~cm$^{-1}$ at low temperature in a uniform way,
as shown in Fig.~\ref{fig:modes}(e), and can be described by the anharmonic decay model,
showing none of the anomalous behavior observed in the position, although there is some
evidence the oscillator strength of this mode may increase slightly at low temperature,
shown in Fig.~\ref{fig:modes}(f).

The optical conductivity the fits were performed on have a typical wave number resolution
of 1.8~cm$^{-1}$.  Measurements with a resolution of 0.2~cm$^{-1}$ were performed in the
far-infrared region above and below $T_{\rm N}$ revealed that while the two modes have the
same positions has previously reported, the line width of the low-frequency mode of
$\simeq 0.2$~cm$^{-1}$ suggests it is limited by the resolution of the instrument, while the
high-frequency mode has a width of 0.53~cm$^{-1}$, indicating that this is likely its
intrinsic value \cite{SupMat}.  Interestingly, neither mode splits below $T_{\rm N}$,
suggesting the antiferromagnetic ground state does not result in a significant lattice
distortion.

%
%
\begin{figure}[t]
\centerline{\includegraphics[width=3.2in]{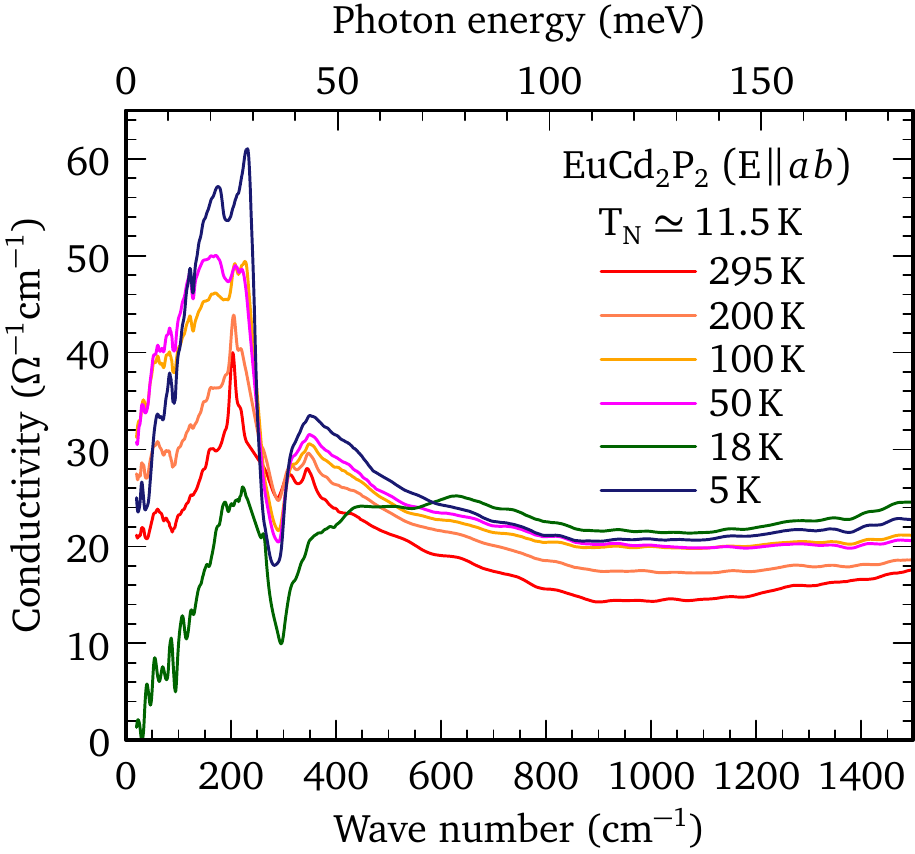}}
\caption{The temperature dependence of the optical conductivity of EuCd$_2$P$_2$ for
light polarized in the \emph{a-b} planes in the far-infrared region with the
infrared-active $E_u$ modes removed, revealing the non-Drude response.}
\label{fig:remove}
\end{figure}

%
%
\subsection{Electronic response}
The determination of the vibrational parameters for the two $E_u$ modes allows these
features to be subtracted from the optical conductivity in Fig.~\ref{fig:sigma},
resulting in the residual conductivity shown in Fig.~\ref{fig:remove}.  There are
several things about the residual conductivity that are unusual.  The first is
that the conductivity is non-Drude; it is initially increasing as a function of
frequency, reaching a maximum at $\simeq 200$~cm$^{-1}$ before encountering a
notch-like feature between $250 - 300$~cm$^{-1}$.
The Drude form for the real part of the optical conductivity may be written as
\begin{equation}
  \sigma_1(\omega) = \left( \frac{2\pi}{Z_0}\right)
  \frac{\omega_{p,D}^2\tau_D}{1+\omega^2\tau_D^2},
\end{equation}
which has the form of a Lorentzian centered at zero frequency with a width  of
$1/\tau_D$.

%
%
\begin{figure}[tb]
\centerline{\includegraphics[width=3.1in]{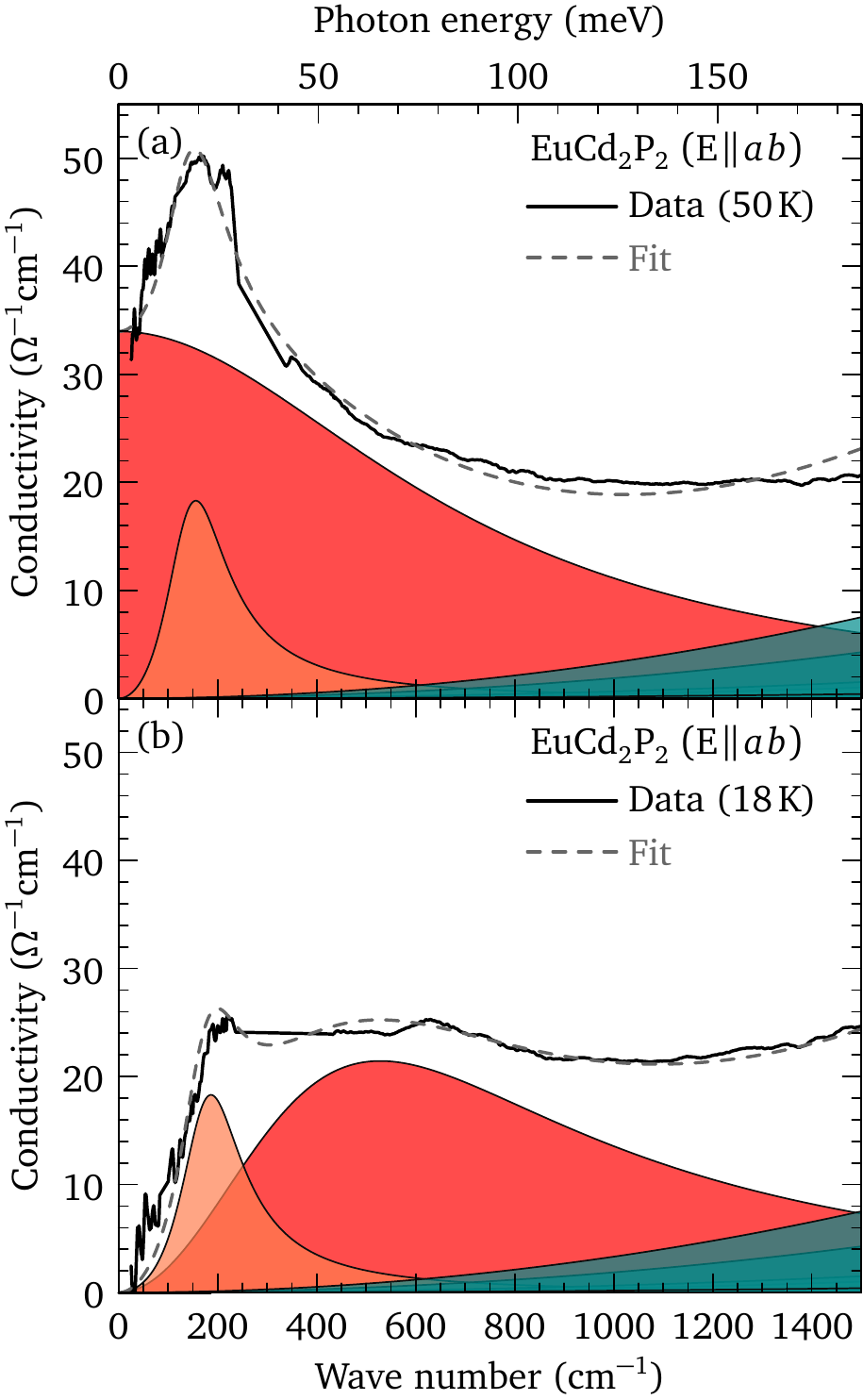}}
\caption{The fit using the Drude-Lorentz model to the real part of the optical conductivity
of EuCd$_2$P$_2$ for light polarized in the \emph{a-b} planes with the infrared-active $E_u$
modes and the notch-like feature removed at (a) 50~K, and (b) 18~K.  The fit has been decomposed
into the contributions for the free-carrier response as well as the bound excitations.  The
spectral weight associated with the Drude component is observed to shift to a bound excitation
(red curve) with a weaker component that does not change, while the darker colors indicate the
low-frequency tails of the mid-infrared absorptions.}
\label{fig:fits}
\end{figure}

The origin of the notch-like feature may arise from the terraced nature of the
crystal surface, which can introduce structure from the \emph{c} axis into the
\emph{ab}-plane optical properties. Alternatively, it has been established that in
anisotropic media such as the cuprate materials, measuring the reflectivity at
anything other than a normal angle of incidence, or having a slightly misoriented
surface, can allow \emph{c}-axis longitudinal optic (LO) modes to manifest
themselves as antiresonaces (resonances) in the metallic (insulating) electronic
background \cite{Berreman1970,Reedyk1992,Homes2007}.  The results from an \emph{a-c}
[(101)] face allows the positions of the $A_{2u}$ transverse optic (TO) modes to be
determined and the lower bound of the LO modes to be estimated; comparison with
the in-plane results reveal that the artifacts in the \emph{a-b} plane reflectivity
arise from the \emph{c} axis TO modes.  The terraces lead to a mixing of the in-plane
and \emph{c} axis reflectivity, thus the notch-like feature is considered to be an
artifact and not intrinsic and will be ignored in fits to the residual conductivity;
this is discussed in detail in the Supplementary Materials \cite{SupMat}.

%
%
The residual optical conductivity has been fit using the Drude-Lorentz model
and the results for two representative fits at 50 and 18~K are shown in
Figs.~\ref{fig:fits}(a) and \ref{fig:fits}(b), respectively; the bound excitations
associated with the interband transitions have been held fixed, partly to allow for
a more reliable convergence, and also because little temperature dependence of these
features is expected (Fig.~\ref{fig:inter}).  The quality of the fits
are quite good.  In the metallic state at 50~K, the Drude response dominates the
free-carrier response with $\omega_{p,D} \simeq 1200$~cm$^{-1}$ with a scattering rate
of $1/\tau_D\simeq 700$~cm$^{-1}$, values that are representative of a marginally-metallic
material.  A weak bound excitation at $\simeq 180$~cm$^{-1}$ has been included to reproduce
the non-Drude response at low frequency; however, it represents only about 10\% of the
spectral weight associated with the free carriers.
At 18~K, the Drude response vanishes and essentially all of the spectral weight associated
with free carriers has been transferred to a bound excitation at $\omega_0\simeq 500$~cm$^{-1}$,
with a width $\gamma_0\simeq 1200$~cm$^{-1}$ and $\Omega_0\simeq 1200$~cm$^{-1}$. The results
of the fits have been summarized in Table~\ref{tab:fits}.
While position of the excitation at $\simeq 180$~cm$^{-1}$ shows little temperature dependence,
it does increase somewhat in strength below $T_{\rm N}$, suggesting that it is affected
by the formation of magnetic order; however, its origin remains unclear.

%
%
\begin{table}[tb]
\caption{The fitted values for the residual optical conductivity of EuCd$_2$P$_2$
for light polarized in the \emph{a-b} planes, consisting of Drude parameters in
the metallic state, and a bound excitation in the semiconducting state; the estimated
error on the fitted quantities is approximately 10\%.  All units are in cm$^{-1}$,
unless otherwise indicated.}
\begin{ruledtabular}
\begin{tabular}{c cc ccc}
  $T\,({\rm K})$  & $\omega_{p,D}$ & $1/\tau_D$ & $\omega_0$ & $\gamma_0$ & $\Omega_0$ \\
\cline{1-6}
  295 &   993 &  756 & & & \\
  200 &  1100 &  753 & & & \\
  100 &  1160 &  680 & & & \\
   50 &  1193 &  698 & & & \\
   25 &  1311 & 1265 & & & \\
   18 &       &      & 486 & 1180 & 1183 \\
   15 &       &      & 539 & 1028 & 1154 \\
   10 &       &      & 426 &  770 & 1078 \\
    5 &  1170 &  919 & & & \\
\end{tabular}
\end{ruledtabular}
\footnotetext[1] {A low-frequency bound excitation is also included at all temperatures to describe
the low-frequency conductivity; while it's parameters vary, $\omega_0 \approx \gamma_0 \simeq 180$~cm$^{-1}$,
and $\Omega_0\simeq 400$~cm$^{-1}$, except at 5~K where it increases to $\simeq 600$~cm$^{-1}$.}
\label{tab:fits}
\end{table}

The temperature dependence of Drude plasma frequency is roughly constant with a value
of $\omega_{p,D} \simeq 1150\pm 150$~cm$^{-1}$.  When the Drude term vanishes at
temperatures close to the resistivity maximum, the free-carrier spectral weight is
transferred into a localized excitation with $\Omega_0 \simeq \omega_{p,D}$, which is
also temperature independent.  On the other hand, the Drude scattering rate decreases
from $1/\tau_D \simeq 760$ to 700~cm$^{-1}$ at 50~K, but then increases dramatically
at 25~K, just above the resistivity maximum, to $1/\tau_D \simeq 1270$~cm$^{-1}$.
%
%
It is tempting to associate the rapid increase in $1/\tau_D$ close to $T_{\rm N}$
with the scattering of carriers from antiferromagnetic fluctuations.  This argument
would be reasonable if the maximum in the resistivity occurred just above $T_{\rm N}$;
however, it occurs at $\simeq 18$~K, nearly twice $T_{\rm N}$.  Moreover, the resistivity
in this material increases by roughly two orders of magnitude, far greater than the modest
increases that are observed at $\simeq T_{\rm N}$ in the As and Sb compounds.

%
%
A more conventional approach relies on polaron effects to describe the transport behavior
in the colossal magnetoresistance materials \cite{Emin1993,Alexandrov-Book,Millis1996,
Roder1996,Lee1997,Salamon2001}.
For instance, a transition from large (delocalized carriers) to small polarons (localized
carriers) has been used to explain the metal-insulator transition in manganites
\cite{Lanzara1998,Hartinger2004}.  The application of polarons to this material is
attractive because the shape of the electronic absorption is that of an asymmetric
Gaussian with a long high-frequency tail \cite{Hartinger2004}, which it could be argued
better describes the electronic background in Fig.~\ref{fig:fits}(b) than the two
Lorentzian oscillators; additionally, the peak of the localized carriers occurs at
$\simeq 480\pm 50$~cm$^{-1}$, which is roughly twice the energy of the high-frequency
$E_u$ mode, consistent with a polaronic mechanism.  There is are several problems with
this interpretation.  The first is that small lattice polarons are expected to form in
systems with strong electron-phonon coupling \cite{Salamon2001}; however, as previously
noted, the narrow line shapes of the infrared-active $E_u$ modes do not support the
notion of strong electron-phonon coupling.  Additionally, the size of this effect
is orders of magnitude smaller than what is observed in the manganites, and finally,
no removal of degeneracy due to the coupling of the infrared-active modes to a lattice
distortion is observed.  Consequently, the polaronic view is not favored.

%
%
The most compelling explanation for the dramatic increase of the resistivity above
$T_{\rm N}$ lies in the recent observation of the formation of ferromagnetism at
$\simeq 2T_{\rm N}$ \cite{Sunko2022}; ferromagnetic clusters form in a paramagnetic
background, resulting in carriers becoming localized into spin-polarized clusters
due to spin-carrier coupling.  Optically, this is observed as the transfer of
spectral weight from the free-carriers into a localization peak.  The resistivity
continues to increase until the ferromagnetic regions begin to merge into a
contiguous network just above $T_{\rm N}$, at which point the resistivity begins
to decrease and the spectral weight is transferred back into the free-carrier component.
The ferromagnetic regions are observed to persist below $T_{\rm N}$ in the metallic
antiferromagnetic phase where the conductivity has improved slightly, likely due
to a decrease in spin fluctuations \cite{Zhang2010}.

%
\section{Conclusions}
The optical properties of a single crystal of EuCd$_2$P$_2$ have been determined for
light polarized in the \emph{a-b} planes above and below $T_{\rm N}$ over a wide
frequency range.  At room temperature, the real part of the optical conductivity
reveals a marginally-metallic material, consisting of a weak free-carrier component
with the onset of interband transitions above $\simeq 2000$~cm$^{-1}$.  Two sharp
infrared-active $E_u$ modes are observed at $\simeq 89$ and 239~cm$^{-1}$.  In
addition, a spurious notch-like feature observed in the in-plane conductivity is
attributed to \emph{c}-axis contamination due to terraces in the crystal surface
\cite{SupMat}.  As the temperature is lowered, there is a modest increase in the
low-frequency conductivity; however, below about 50~K the conductivity begins to
decrease until at $\simeq 18$~K (just below $2T_{\rm N}$), a dramatic change is
observed as the majority of the free-carriers enter into a localized state, before
reverting back to metallic behavior again below $T_{\rm N}$.  The loss and
subsequent restoration of the free-carrier electronic background has a relatively
minor effect on the nature of the $E_u$ modes; however, it results in a dramatic
change in the line shapes of these vibrations in the reflectivity \cite{SupMat}.
The localization of the free carriers is intimately connected with the magnetism
in this material.  While several scenarios are considered, the prevailing
explanation for the resistivity maximum and carrier localization is the formation
of ferromagnetic domains below $\simeq 2T_{\rm N}$ that result in spin-polarized
clusters due to spin-carrier coupling; once these domains form a contiguous network,
the resistivity decreases and the free-carrier component is restored \cite{Sunko2022}.

%
%
%
\begin{acknowledgments}
We would like to acknowledge useful discussions with A. Akrap, V. Sunko and
J. Orenstein.
The work at Boston College was funded by the National Science Foundation
under Award No. NSF/DMR-1708929.
Work at Brookhaven National Laboratory was supported by the Office of Science, U.S.
Department of Energy under Contract No. DE-SC0012704.
\end{acknowledgments}

%
%
%

\begin{thebibliography}{38}%
\makeatletter
\providecommand \@ifxundefined [1]{%
 \@ifx{#1\undefined}
}%
\providecommand \@ifnum [1]{%
 \ifnum #1\expandafter \@firstoftwo
 \else \expandafter \@secondoftwo
 \fi
}%
\providecommand \@ifx [1]{%
 \ifx #1\expandafter \@firstoftwo
 \else \expandafter \@secondoftwo
 \fi
}%
\providecommand \natexlab [1]{#1}%
\providecommand \enquote  [1]{``#1''}%
\providecommand \bibnamefont  [1]{#1}%
\providecommand \bibfnamefont [1]{#1}%
\providecommand \citenamefont [1]{#1}%
\providecommand \href@noop [0]{\@secondoftwo}%
\providecommand \href [0]{\begingroup \@sanitize@url \@href}%
\providecommand \@href[1]{\@@startlink{#1}\@@href}%
\providecommand \@@href[1]{\endgroup#1\@@endlink}%
\providecommand \@sanitize@url [0]{\catcode `\\12\catcode `\$12\catcode
  `\&12\catcode `\#12\catcode `\^12\catcode `\_12\catcode `\%12\relax}%
\providecommand \@@startlink[1]{}%
\providecommand \@@endlink[0]{}%
\providecommand \url  [0]{\begingroup\@sanitize@url \@url }%
\providecommand \@url [1]{\endgroup\@href {#1}{\urlprefix }}%
\providecommand \urlprefix  [0]{URL }%
\providecommand \Eprint [0]{\href }%
\providecommand \doibase [0]{https://doi.org/}%
\providecommand \selectlanguage [0]{\@gobble}%
\providecommand \bibinfo  [0]{\@secondoftwo}%
\providecommand \bibfield  [0]{\@secondoftwo}%
\providecommand \translation [1]{[#1]}%
\providecommand \BibitemOpen [0]{}%
\providecommand \bibitemStop [0]{}%
\providecommand \bibitemNoStop [0]{.\EOS\space}%
\providecommand \EOS [0]{\spacefactor3000\relax}%
\providecommand \BibitemShut  [1]{\csname bibitem#1\endcsname}%
\let\auto@bib@innerbib\@empty
\bibitem [{\citenamefont {Sunko}\ \emph {et~al.}(2022)\citenamefont {Sunko},
  \citenamefont {Vranas}, \citenamefont {Homes}, \citenamefont {Lee},
  \citenamefont {Donoway}, \citenamefont {Wang}, \citenamefont {Balguri},
  \citenamefont {Mahendru}, \citenamefont {Ruiz}, \citenamefont {Gunn},
  \citenamefont {Basak}, \citenamefont {Schierle}, \citenamefont {Weschke},
  \citenamefont {Tafti}, \citenamefont {Frano},\ and\ \citenamefont
  {Orenstein}}]{Sunko2022}%
  \BibitemOpen
  \bibfield  {author} {\bibinfo {author} {\bibfnamefont {V.}~\bibnamefont
  {Sunko}}, \bibinfo {author} {\bibfnamefont {Y.~S.~M.}\ \bibnamefont
  {Vranas}}, \bibinfo {author} {\bibfnamefont {C.~C.}\ \bibnamefont {Homes}},
  \bibinfo {author} {\bibfnamefont {C.}~\bibnamefont {Lee}}, \bibinfo {author}
  {\bibfnamefont {E.}~\bibnamefont {Donoway}}, \bibinfo {author} {\bibfnamefont
  {Z.-C.}\ \bibnamefont {Wang}}, \bibinfo {author} {\bibfnamefont
  {S.}~\bibnamefont {Balguri}}, \bibinfo {author} {\bibfnamefont {M.~B.}\
  \bibnamefont {Mahendru}}, \bibinfo {author} {\bibfnamefont {A.}~\bibnamefont
  {Ruiz}}, \bibinfo {author} {\bibfnamefont {B.}~\bibnamefont {Gunn}}, \bibinfo
  {author} {\bibfnamefont {R.}~\bibnamefont {Basak}}, \bibinfo {author}
  {\bibfnamefont {E.}~\bibnamefont {Schierle}}, \bibinfo {author}
  {\bibfnamefont {E.}~\bibnamefont {Weschke}}, \bibinfo {author} {\bibfnamefont
  {F.}~\bibnamefont {Tafti}}, \bibinfo {author} {\bibfnamefont
  {A.}~\bibnamefont {Frano}},\ and\ \bibinfo {author} {\bibfnamefont
  {J.}~\bibnamefont {Orenstein}},\ }\bibfield  {title} {\bibinfo {title}
  {{Spin-carrier coupling induced ferromagnetism and giant resistivity peak in
  EuCd$_2$P$_2$}},\ }\Eprint {https://arxiv.org/abs/2208.05499v1}
  {arXiv:2208.05499v1}  (\bibinfo {year} {2022})\BibitemShut {NoStop}%
\bibitem [{\citenamefont {Wang}\ \emph {et~al.}(2016)\citenamefont {Wang},
  \citenamefont {Wu}, \citenamefont {Shi},\ and\ \citenamefont
  {Wang}}]{Wang2016}%
  \BibitemOpen
  \bibfield  {author} {\bibinfo {author} {\bibfnamefont {H.~P.}\ \bibnamefont
  {Wang}}, \bibinfo {author} {\bibfnamefont {D.~S.}\ \bibnamefont {Wu}},
  \bibinfo {author} {\bibfnamefont {Y.~G.}\ \bibnamefont {Shi}},\ and\ \bibinfo
  {author} {\bibfnamefont {N.~L.}\ \bibnamefont {Wang}},\ }\bibfield  {title}
  {\bibinfo {title} {{Anisotropic transport and optical spectroscopy study on
  antiferromagnetic triangular lattice EuCd$_{2}$As$_{2}$: An interplay between
  magnetism and charge transport properties}},\ }\href
  {https://doi.org/10.1103/PhysRevB.94.045112} {\bibfield  {journal} {\bibinfo
  {journal} {Phys. Rev. B}\ }\textbf {\bibinfo {volume} {94}},\ \bibinfo
  {pages} {045112} (\bibinfo {year} {2016})}\BibitemShut {NoStop}%
\bibitem [{\citenamefont {Ma}\ \emph {et~al.}(2019)\citenamefont {Ma},
  \citenamefont {Nie}, \citenamefont {Yi}, \citenamefont {Jandke},
  \citenamefont {Shang}, \citenamefont {Yao}, \citenamefont {Naamneh},
  \citenamefont {Yan}, \citenamefont {Sun}, \citenamefont {Chikina},
  \citenamefont {Strocov}, \citenamefont {Medarde}, \citenamefont {Song},
  \citenamefont {Xiong}, \citenamefont {Xu}, \citenamefont {Wulfhekel},
  \citenamefont {Mesot}, \citenamefont {Reticcioli}, \citenamefont {Franchini},
  \citenamefont {Mudry}, \citenamefont {M\"{u}ller}, \citenamefont {Shi},
  \citenamefont {Qian}, \citenamefont {Ding},\ and\ \citenamefont
  {Shi}}]{Ma2019}%
  \BibitemOpen
  \bibfield  {author} {\bibinfo {author} {\bibfnamefont {J.-Z.}\ \bibnamefont
  {Ma}}, \bibinfo {author} {\bibfnamefont {S.~M.}\ \bibnamefont {Nie}},
  \bibinfo {author} {\bibfnamefont {C.~J.}\ \bibnamefont {Yi}}, \bibinfo
  {author} {\bibfnamefont {J.}~\bibnamefont {Jandke}}, \bibinfo {author}
  {\bibfnamefont {T.}~\bibnamefont {Shang}}, \bibinfo {author} {\bibfnamefont
  {M.~Y.}\ \bibnamefont {Yao}}, \bibinfo {author} {\bibfnamefont
  {M.}~\bibnamefont {Naamneh}}, \bibinfo {author} {\bibfnamefont {L.~Q.}\
  \bibnamefont {Yan}}, \bibinfo {author} {\bibfnamefont {Y.}~\bibnamefont
  {Sun}}, \bibinfo {author} {\bibfnamefont {A.}~\bibnamefont {Chikina}},
  \bibinfo {author} {\bibfnamefont {V.~N.}\ \bibnamefont {Strocov}}, \bibinfo
  {author} {\bibfnamefont {M.}~\bibnamefont {Medarde}}, \bibinfo {author}
  {\bibfnamefont {M.}~\bibnamefont {Song}}, \bibinfo {author} {\bibfnamefont
  {Y.-M.}\ \bibnamefont {Xiong}}, \bibinfo {author} {\bibfnamefont
  {G.}~\bibnamefont {Xu}}, \bibinfo {author} {\bibfnamefont {W.}~\bibnamefont
  {Wulfhekel}}, \bibinfo {author} {\bibfnamefont {J.}~\bibnamefont {Mesot}},
  \bibinfo {author} {\bibfnamefont {M.}~\bibnamefont {Reticcioli}}, \bibinfo
  {author} {\bibfnamefont {C.}~\bibnamefont {Franchini}}, \bibinfo {author}
  {\bibfnamefont {C.}~\bibnamefont {Mudry}}, \bibinfo {author} {\bibfnamefont
  {M.}~\bibnamefont {M\"{u}ller}}, \bibinfo {author} {\bibfnamefont {Y.~G.}\
  \bibnamefont {Shi}}, \bibinfo {author} {\bibfnamefont {T.}~\bibnamefont
  {Qian}}, \bibinfo {author} {\bibfnamefont {H.}~\bibnamefont {Ding}},\ and\
  \bibinfo {author} {\bibfnamefont {M.}~\bibnamefont {Shi}},\ }\bibfield
  {title} {\bibinfo {title} {{Spin fluctuation induced Weyl semimetal state in
  the paramagnetic phase of EuCd$_2$As$_2$}},\ }\href
  {https://doi.org/10.1126/sciadv.aaw4718} {\bibfield  {journal} {\bibinfo
  {journal} {Sci. Adv.}\ }\textbf {\bibinfo {volume} {5}},\ \bibinfo {pages}
  {eaaw4718} (\bibinfo {year} {2019})}\BibitemShut {NoStop}%
\bibitem [{\citenamefont {Wang}\ \emph {et~al.}(2019)\citenamefont {Wang},
  \citenamefont {Jo}, \citenamefont {Kuthanazhi}, \citenamefont {Wu},
  \citenamefont {McQueeney}, \citenamefont {Kaminski},\ and\ \citenamefont
  {Canfield}}]{Wang2019}%
  \BibitemOpen
  \bibfield  {author} {\bibinfo {author} {\bibfnamefont {L.-L.}\ \bibnamefont
  {Wang}}, \bibinfo {author} {\bibfnamefont {N.~H.}\ \bibnamefont {Jo}},
  \bibinfo {author} {\bibfnamefont {B.}~\bibnamefont {Kuthanazhi}}, \bibinfo
  {author} {\bibfnamefont {Y.}~\bibnamefont {Wu}}, \bibinfo {author}
  {\bibfnamefont {R.~J.}\ \bibnamefont {McQueeney}}, \bibinfo {author}
  {\bibfnamefont {A.}~\bibnamefont {Kaminski}},\ and\ \bibinfo {author}
  {\bibfnamefont {P.~C.}\ \bibnamefont {Canfield}},\ }\bibfield  {title}
  {\bibinfo {title} {{Single pair of Weyl fermions in the half-metallic
  semimetal EuCd$_2$As$_2$}},\ }\href
  {https://doi.org/10.1103/PhysRevB.99.245147} {\bibfield  {journal} {\bibinfo
  {journal} {Phys. Rev. B}\ }\textbf {\bibinfo {volume} {99}},\ \bibinfo
  {pages} {245147} (\bibinfo {year} {2019})}\BibitemShut {NoStop}%
\bibitem [{\citenamefont {Behrends}\ \emph {et~al.}(2019)\citenamefont
  {Behrends}, \citenamefont {Ilan},\ and\ \citenamefont
  {Bardarson}}]{Behrends2019}%
  \BibitemOpen
  \bibfield  {author} {\bibinfo {author} {\bibfnamefont {J.}~\bibnamefont
  {Behrends}}, \bibinfo {author} {\bibfnamefont {R.}~\bibnamefont {Ilan}},\
  and\ \bibinfo {author} {\bibfnamefont {J.~H.}\ \bibnamefont {Bardarson}},\
  }\bibfield  {title} {\bibinfo {title} {{Anomalous conductance scaling in
  strained Weyl semimetals}},\ }\href
  {https://doi.org/10.1103/PhysRevResearch.1.032028} {\bibfield  {journal}
  {\bibinfo  {journal} {Phys. Rev. Research}\ }\textbf {\bibinfo {volume}
  {1}},\ \bibinfo {pages} {032028} (\bibinfo {year} {2019})}\BibitemShut
  {NoStop}%
\bibitem [{\citenamefont {Soh}\ \emph {et~al.}(2019)\citenamefont {Soh},
  \citenamefont {de~Juan}, \citenamefont {Vergniory}, \citenamefont
  {Schr\"oter}, \citenamefont {Rahn}, \citenamefont {Yan}, \citenamefont
  {Jiang}, \citenamefont {Bristow}, \citenamefont {Reiss}, \citenamefont
  {Blandy}, \citenamefont {Guo}, \citenamefont {Shi}, \citenamefont {Kim},
  \citenamefont {McCollam}, \citenamefont {Simon}, \citenamefont {Chen},
  \citenamefont {Coldea},\ and\ \citenamefont {Boothroyd}}]{Soh2019}%
  \BibitemOpen
  \bibfield  {author} {\bibinfo {author} {\bibfnamefont {J.-R.}\ \bibnamefont
  {Soh}}, \bibinfo {author} {\bibfnamefont {F.}~\bibnamefont {de~Juan}},
  \bibinfo {author} {\bibfnamefont {M.~G.}\ \bibnamefont {Vergniory}}, \bibinfo
  {author} {\bibfnamefont {N.~B.~M.}\ \bibnamefont {Schr\"oter}}, \bibinfo
  {author} {\bibfnamefont {M.~C.}\ \bibnamefont {Rahn}}, \bibinfo {author}
  {\bibfnamefont {D.~Y.}\ \bibnamefont {Yan}}, \bibinfo {author} {\bibfnamefont
  {J.}~\bibnamefont {Jiang}}, \bibinfo {author} {\bibfnamefont
  {M.}~\bibnamefont {Bristow}}, \bibinfo {author} {\bibfnamefont
  {P.}~\bibnamefont {Reiss}}, \bibinfo {author} {\bibfnamefont {J.~N.}\
  \bibnamefont {Blandy}}, \bibinfo {author} {\bibfnamefont {Y.~F.}\
  \bibnamefont {Guo}}, \bibinfo {author} {\bibfnamefont {Y.~G.}\ \bibnamefont
  {Shi}}, \bibinfo {author} {\bibfnamefont {T.~K.}\ \bibnamefont {Kim}},
  \bibinfo {author} {\bibfnamefont {A.}~\bibnamefont {McCollam}}, \bibinfo
  {author} {\bibfnamefont {S.~H.}\ \bibnamefont {Simon}}, \bibinfo {author}
  {\bibfnamefont {Y.}~\bibnamefont {Chen}}, \bibinfo {author} {\bibfnamefont
  {A.~I.}\ \bibnamefont {Coldea}},\ and\ \bibinfo {author} {\bibfnamefont
  {A.~T.}\ \bibnamefont {Boothroyd}},\ }\bibfield  {title} {\bibinfo {title}
  {{Ideal Weyl semimetal induced by magnetic exchange}},\ }\href
  {https://doi.org/10.1103/PhysRevB.100.201102} {\bibfield  {journal} {\bibinfo
   {journal} {Phys. Rev. B}\ }\textbf {\bibinfo {volume} {100}},\ \bibinfo
  {pages} {201102} (\bibinfo {year} {2019})}\BibitemShut {NoStop}%
\bibitem [{\citenamefont {Jo}\ \emph {et~al.}(2020)\citenamefont {Jo},
  \citenamefont {Kuthanazhi}, \citenamefont {Wu}, \citenamefont {Timmons},
  \citenamefont {Kim}, \citenamefont {Zhou}, \citenamefont {Wang},
  \citenamefont {Ueland}, \citenamefont {Palasyuk}, \citenamefont {Ryan},
  \citenamefont {McQueeney}, \citenamefont {Lee}, \citenamefont {Schrunk},
  \citenamefont {Burkov}, \citenamefont {Prozorov}, \citenamefont {Bud'ko},
  \citenamefont {Kaminski},\ and\ \citenamefont {Canfield}}]{Jo2020}%
  \BibitemOpen
  \bibfield  {author} {\bibinfo {author} {\bibfnamefont {N.~H.}\ \bibnamefont
  {Jo}}, \bibinfo {author} {\bibfnamefont {B.}~\bibnamefont {Kuthanazhi}},
  \bibinfo {author} {\bibfnamefont {Y.}~\bibnamefont {Wu}}, \bibinfo {author}
  {\bibfnamefont {E.}~\bibnamefont {Timmons}}, \bibinfo {author} {\bibfnamefont
  {T.-H.}\ \bibnamefont {Kim}}, \bibinfo {author} {\bibfnamefont
  {L.}~\bibnamefont {Zhou}}, \bibinfo {author} {\bibfnamefont {L.-L.}\
  \bibnamefont {Wang}}, \bibinfo {author} {\bibfnamefont {B.~G.}\ \bibnamefont
  {Ueland}}, \bibinfo {author} {\bibfnamefont {A.}~\bibnamefont {Palasyuk}},
  \bibinfo {author} {\bibfnamefont {D.~H.}\ \bibnamefont {Ryan}}, \bibinfo
  {author} {\bibfnamefont {R.~J.}\ \bibnamefont {McQueeney}}, \bibinfo {author}
  {\bibfnamefont {K.}~\bibnamefont {Lee}}, \bibinfo {author} {\bibfnamefont
  {B.}~\bibnamefont {Schrunk}}, \bibinfo {author} {\bibfnamefont {A.~A.}\
  \bibnamefont {Burkov}}, \bibinfo {author} {\bibfnamefont {R.}~\bibnamefont
  {Prozorov}}, \bibinfo {author} {\bibfnamefont {S.~L.}\ \bibnamefont
  {Bud'ko}}, \bibinfo {author} {\bibfnamefont {A.}~\bibnamefont {Kaminski}},\
  and\ \bibinfo {author} {\bibfnamefont {P.~C.}\ \bibnamefont {Canfield}},\
  }\bibfield  {title} {\bibinfo {title} {Manipulating magnetism in the
  topological semimetal {EuCd}$_{2}${As}$_{2}$},\ }\href
  {https://doi.org/10.1103/PhysRevB.101.140402} {\bibfield  {journal} {\bibinfo
   {journal} {Phys. Rev. B}\ }\textbf {\bibinfo {volume} {101}},\ \bibinfo
  {pages} {140402(R)} (\bibinfo {year} {2020})}\BibitemShut {NoStop}%
\bibitem [{\citenamefont {Su}\ \emph {et~al.}(2020)\citenamefont {Su},
  \citenamefont {Gong}, \citenamefont {Shi}, \citenamefont {Yang},
  \citenamefont {Wang}, \citenamefont {Xia}, \citenamefont {Yu}, \citenamefont
  {Guo}, \citenamefont {Wang}, \citenamefont {Ding}, \citenamefont {Xu},
  \citenamefont {Li}, \citenamefont {Wang}, \citenamefont {Zou}, \citenamefont
  {Yu}, \citenamefont {Zhu}, \citenamefont {Chen}, \citenamefont {Liu},
  \citenamefont {Liu}, \citenamefont {Li},\ and\ \citenamefont {Guo}}]{Su2020}%
  \BibitemOpen
  \bibfield  {author} {\bibinfo {author} {\bibfnamefont {H.}~\bibnamefont
  {Su}}, \bibinfo {author} {\bibfnamefont {B.}~\bibnamefont {Gong}}, \bibinfo
  {author} {\bibfnamefont {W.}~\bibnamefont {Shi}}, \bibinfo {author}
  {\bibfnamefont {H.}~\bibnamefont {Yang}}, \bibinfo {author} {\bibfnamefont
  {H.}~\bibnamefont {Wang}}, \bibinfo {author} {\bibfnamefont {W.}~\bibnamefont
  {Xia}}, \bibinfo {author} {\bibfnamefont {Z.}~\bibnamefont {Yu}}, \bibinfo
  {author} {\bibfnamefont {P.-J.}\ \bibnamefont {Guo}}, \bibinfo {author}
  {\bibfnamefont {J.}~\bibnamefont {Wang}}, \bibinfo {author} {\bibfnamefont
  {L.}~\bibnamefont {Ding}}, \bibinfo {author} {\bibfnamefont {L.}~\bibnamefont
  {Xu}}, \bibinfo {author} {\bibfnamefont {X.}~\bibnamefont {Li}}, \bibinfo
  {author} {\bibfnamefont {X.}~\bibnamefont {Wang}}, \bibinfo {author}
  {\bibfnamefont {Z.}~\bibnamefont {Zou}}, \bibinfo {author} {\bibfnamefont
  {N.}~\bibnamefont {Yu}}, \bibinfo {author} {\bibfnamefont {Z.}~\bibnamefont
  {Zhu}}, \bibinfo {author} {\bibfnamefont {Y.}~\bibnamefont {Chen}}, \bibinfo
  {author} {\bibfnamefont {Z.}~\bibnamefont {Liu}}, \bibinfo {author}
  {\bibfnamefont {K.}~\bibnamefont {Liu}}, \bibinfo {author} {\bibfnamefont
  {G.}~\bibnamefont {Li}},\ and\ \bibinfo {author} {\bibfnamefont
  {Y.}~\bibnamefont {Guo}},\ }\bibfield  {title} {\bibinfo {title} {{Magnetic
  exchange induced Weyl state in a semimetal EuCd$_2$Sb$_2$}},\ }\href
  {https://doi.org/10.1063/1.5129467} {\bibfield  {journal} {\bibinfo
  {journal} {APL Materials}\ }\textbf {\bibinfo {volume} {8}},\ \bibinfo
  {pages} {011109} (\bibinfo {year} {2020})}\BibitemShut {NoStop}%
\bibitem [{\citenamefont {Ma}\ \emph {et~al.}(2020)\citenamefont {Ma},
  \citenamefont {Wang}, \citenamefont {Nie}, \citenamefont {Yi}, \citenamefont
  {Xu}, \citenamefont {Li}, \citenamefont {Jandke}, \citenamefont {Wulfhekel},
  \citenamefont {Huang}, \citenamefont {West}, \citenamefont {Richard},
  \citenamefont {Chikina}, \citenamefont {Strocov}, \citenamefont {Mesot},
  \citenamefont {Weng}, \citenamefont {Zhang}, \citenamefont {Shi},
  \citenamefont {Qian}, \citenamefont {Shi},\ and\ \citenamefont
  {Ding}}]{Ma2020}%
  \BibitemOpen
  \bibfield  {author} {\bibinfo {author} {\bibfnamefont {J.}~\bibnamefont
  {Ma}}, \bibinfo {author} {\bibfnamefont {H.}~\bibnamefont {Wang}}, \bibinfo
  {author} {\bibfnamefont {S.}~\bibnamefont {Nie}}, \bibinfo {author}
  {\bibfnamefont {C.}~\bibnamefont {Yi}}, \bibinfo {author} {\bibfnamefont
  {Y.}~\bibnamefont {Xu}}, \bibinfo {author} {\bibfnamefont {H.}~\bibnamefont
  {Li}}, \bibinfo {author} {\bibfnamefont {J.}~\bibnamefont {Jandke}}, \bibinfo
  {author} {\bibfnamefont {W.}~\bibnamefont {Wulfhekel}}, \bibinfo {author}
  {\bibfnamefont {Y.}~\bibnamefont {Huang}}, \bibinfo {author} {\bibfnamefont
  {D.}~\bibnamefont {West}}, \bibinfo {author} {\bibfnamefont {P.}~\bibnamefont
  {Richard}}, \bibinfo {author} {\bibfnamefont {A.}~\bibnamefont {Chikina}},
  \bibinfo {author} {\bibfnamefont {V.~N.}\ \bibnamefont {Strocov}}, \bibinfo
  {author} {\bibfnamefont {J.}~\bibnamefont {Mesot}}, \bibinfo {author}
  {\bibfnamefont {H.}~\bibnamefont {Weng}}, \bibinfo {author} {\bibfnamefont
  {S.}~\bibnamefont {Zhang}}, \bibinfo {author} {\bibfnamefont
  {Y.}~\bibnamefont {Shi}}, \bibinfo {author} {\bibfnamefont {T.}~\bibnamefont
  {Qian}}, \bibinfo {author} {\bibfnamefont {M.}~\bibnamefont {Shi}},\ and\
  \bibinfo {author} {\bibfnamefont {H.}~\bibnamefont {Ding}},\ }\bibfield
  {title} {\bibinfo {title} {{Emergence of Nontrivial Low-Energy Dirac Fermions
  in Antiferromagnetic EuCd$_2$As$_2$}},\ }\href
  {https://doi.org/https://doi.org/10.1002/adma.201907565} {\bibfield
  {journal} {\bibinfo  {journal} {Adv. Mater.}\ }\textbf {\bibinfo {volume}
  {32}},\ \bibinfo {pages} {1907565} (\bibinfo {year} {2020})}\BibitemShut
  {NoStop}%
\bibitem [{\citenamefont {Xu}\ \emph {et~al.}(2021)\citenamefont {Xu},
  \citenamefont {Das}, \citenamefont {Ma}, \citenamefont {Yi}, \citenamefont
  {Nie}, \citenamefont {Shi}, \citenamefont {Tiwari}, \citenamefont {Tsirkin},
  \citenamefont {Neupert}, \citenamefont {Medarde}, \citenamefont {Shi},
  \citenamefont {Chang},\ and\ \citenamefont {Shang}}]{Xu2021}%
  \BibitemOpen
  \bibfield  {author} {\bibinfo {author} {\bibfnamefont {Y.}~\bibnamefont
  {Xu}}, \bibinfo {author} {\bibfnamefont {L.}~\bibnamefont {Das}}, \bibinfo
  {author} {\bibfnamefont {J.~Z.}\ \bibnamefont {Ma}}, \bibinfo {author}
  {\bibfnamefont {C.~J.}\ \bibnamefont {Yi}}, \bibinfo {author} {\bibfnamefont
  {S.~M.}\ \bibnamefont {Nie}}, \bibinfo {author} {\bibfnamefont {Y.~G.}\
  \bibnamefont {Shi}}, \bibinfo {author} {\bibfnamefont {A.}~\bibnamefont
  {Tiwari}}, \bibinfo {author} {\bibfnamefont {S.~S.}\ \bibnamefont {Tsirkin}},
  \bibinfo {author} {\bibfnamefont {T.}~\bibnamefont {Neupert}}, \bibinfo
  {author} {\bibfnamefont {M.}~\bibnamefont {Medarde}}, \bibinfo {author}
  {\bibfnamefont {M.}~\bibnamefont {Shi}}, \bibinfo {author} {\bibfnamefont
  {J.}~\bibnamefont {Chang}},\ and\ \bibinfo {author} {\bibfnamefont
  {T.}~\bibnamefont {Shang}},\ }\bibfield  {title} {\bibinfo {title}
  {{Unconventional Transverse Transport above and below the Magnetic Transition
  Temperature in Weyl Semimetal EuCd$_{2}$As$_{2}$}},\ }\href
  {https://doi.org/10.1103/PhysRevLett.126.076602} {\bibfield  {journal}
  {\bibinfo  {journal} {Phys. Rev. Lett.}\ }\textbf {\bibinfo {volume} {126}},\
  \bibinfo {pages} {076602} (\bibinfo {year} {2021})}\BibitemShut {NoStop}%
\bibitem [{\citenamefont {Schellenberg}\ \emph {et~al.}(2011)\citenamefont
  {Schellenberg}, \citenamefont {Pfannenschmidt}, \citenamefont {Eul},
  \citenamefont {Schwickert},\ and\ \citenamefont
  {P\"{o}ttgen}}]{Schellenberg2011}%
  \BibitemOpen
  \bibfield  {author} {\bibinfo {author} {\bibfnamefont {I.}~\bibnamefont
  {Schellenberg}}, \bibinfo {author} {\bibfnamefont {U.}~\bibnamefont
  {Pfannenschmidt}}, \bibinfo {author} {\bibfnamefont {M.}~\bibnamefont {Eul}},
  \bibinfo {author} {\bibfnamefont {C.}~\bibnamefont {Schwickert}},\ and\
  \bibinfo {author} {\bibfnamefont {R.}~\bibnamefont {P\"{o}ttgen}},\
  }\bibfield  {title} {\bibinfo {title} {{A $^{121}$Sb and $^{151}$Eu
  M\"{o}ssbauer Spectroscopic Investigation of EuCd$_2X_2$ ($X$ = P, As, Sb)
  and YbCd$_2$Sb$_2$}},\ }\href
  {https://doi.org/https://doi.org/10.1002/zaac.201100179} {\bibfield
  {journal} {\bibinfo  {journal} {Z. Anorg. Allg. Chem.}\ }\textbf {\bibinfo
  {volume} {637}},\ \bibinfo {pages} {1863} (\bibinfo {year}
  {2011})}\BibitemShut {NoStop}%
\bibitem [{\citenamefont {Rahn}\ \emph {et~al.}(2018)\citenamefont {Rahn},
  \citenamefont {Soh}, \citenamefont {Francoual}, \citenamefont {Veiga},
  \citenamefont {Strempfer}, \citenamefont {Mardegan}, \citenamefont {Yan},
  \citenamefont {Guo}, \citenamefont {Shi},\ and\ \citenamefont
  {Boothroyd}}]{Rahn2018}%
  \BibitemOpen
  \bibfield  {author} {\bibinfo {author} {\bibfnamefont {M.~C.}\ \bibnamefont
  {Rahn}}, \bibinfo {author} {\bibfnamefont {J.-R.}\ \bibnamefont {Soh}},
  \bibinfo {author} {\bibfnamefont {S.}~\bibnamefont {Francoual}}, \bibinfo
  {author} {\bibfnamefont {L.~S.~I.}\ \bibnamefont {Veiga}}, \bibinfo {author}
  {\bibfnamefont {J.}~\bibnamefont {Strempfer}}, \bibinfo {author}
  {\bibfnamefont {J.}~\bibnamefont {Mardegan}}, \bibinfo {author}
  {\bibfnamefont {D.~Y.}\ \bibnamefont {Yan}}, \bibinfo {author} {\bibfnamefont
  {Y.~F.}\ \bibnamefont {Guo}}, \bibinfo {author} {\bibfnamefont {Y.~G.}\
  \bibnamefont {Shi}},\ and\ \bibinfo {author} {\bibfnamefont {A.~T.}\
  \bibnamefont {Boothroyd}},\ }\bibfield  {title} {\bibinfo {title} {{Coupling
  of magnetic order and charge transport in the candidate Dirac semimetal
  EuCd$_{2}$As$_{2}$}},\ }\href {https://doi.org/10.1103/PhysRevB.97.214422}
  {\bibfield  {journal} {\bibinfo  {journal} {Phys. Rev. B}\ }\textbf {\bibinfo
  {volume} {97}},\ \bibinfo {pages} {214422} (\bibinfo {year}
  {2018})}\BibitemShut {NoStop}%
\bibitem [{\citenamefont {Soh}\ \emph {et~al.}(2020)\citenamefont {Soh},
  \citenamefont {Schierle}, \citenamefont {Yan}, \citenamefont {Su},
  \citenamefont {Prabhakaran}, \citenamefont {Weschke}, \citenamefont {Guo},
  \citenamefont {Shi},\ and\ \citenamefont {Boothroyd}}]{Soh2020}%
  \BibitemOpen
  \bibfield  {author} {\bibinfo {author} {\bibfnamefont {J.-R.}\ \bibnamefont
  {Soh}}, \bibinfo {author} {\bibfnamefont {E.}~\bibnamefont {Schierle}},
  \bibinfo {author} {\bibfnamefont {D.~Y.}\ \bibnamefont {Yan}}, \bibinfo
  {author} {\bibfnamefont {H.}~\bibnamefont {Su}}, \bibinfo {author}
  {\bibfnamefont {D.}~\bibnamefont {Prabhakaran}}, \bibinfo {author}
  {\bibfnamefont {E.}~\bibnamefont {Weschke}}, \bibinfo {author} {\bibfnamefont
  {Y.~F.}\ \bibnamefont {Guo}}, \bibinfo {author} {\bibfnamefont {Y.~G.}\
  \bibnamefont {Shi}},\ and\ \bibinfo {author} {\bibfnamefont {A.~T.}\
  \bibnamefont {Boothroyd}},\ }\bibfield  {title} {\bibinfo {title} {{Resonant
  x-ray scattering study of diffuse magnetic scattering from the topological
  semimetals EuCd$_{2}$As$_{2}$ and EuCd$_{2}$Sb$_{2}$}},\ }\href
  {https://doi.org/10.1103/PhysRevB.102.014408} {\bibfield  {journal} {\bibinfo
   {journal} {Phys. Rev. B}\ }\textbf {\bibinfo {volume} {102}},\ \bibinfo
  {pages} {014408} (\bibinfo {year} {2020})}\BibitemShut {NoStop}%
\bibitem [{\citenamefont {Wang}\ \emph {et~al.}(2021)\citenamefont {Wang},
  \citenamefont {Rogers}, \citenamefont {Yao}, \citenamefont {Nichols},
  \citenamefont {Atay}, \citenamefont {Xu}, \citenamefont {Franklin},
  \citenamefont {Sochnikov}, \citenamefont {Ryan}, \citenamefont {Haskel},\
  and\ \citenamefont {Tafti}}]{Wang2021}%
  \BibitemOpen
  \bibfield  {author} {\bibinfo {author} {\bibfnamefont {Z.-C.}\ \bibnamefont
  {Wang}}, \bibinfo {author} {\bibfnamefont {J.~D.}\ \bibnamefont {Rogers}},
  \bibinfo {author} {\bibfnamefont {X.}~\bibnamefont {Yao}}, \bibinfo {author}
  {\bibfnamefont {R.}~\bibnamefont {Nichols}}, \bibinfo {author} {\bibfnamefont
  {K.}~\bibnamefont {Atay}}, \bibinfo {author} {\bibfnamefont {B.}~\bibnamefont
  {Xu}}, \bibinfo {author} {\bibfnamefont {J.}~\bibnamefont {Franklin}},
  \bibinfo {author} {\bibfnamefont {I.}~\bibnamefont {Sochnikov}}, \bibinfo
  {author} {\bibfnamefont {P.~J.}\ \bibnamefont {Ryan}}, \bibinfo {author}
  {\bibfnamefont {D.}~\bibnamefont {Haskel}},\ and\ \bibinfo {author}
  {\bibfnamefont {F.}~\bibnamefont {Tafti}},\ }\bibfield  {title} {\bibinfo
  {title} {Colossal magnetoresistance without mixed valence in a layered
  phosphide crystal},\ }\href
  {https://doi.org/https://doi.org/10.1002/adma.202005755} {\bibfield
  {journal} {\bibinfo  {journal} {Adv. Mater.}\ }\textbf {\bibinfo {volume}
  {33}},\ \bibinfo {pages} {2005755} (\bibinfo {year} {2021})}\BibitemShut
  {NoStop}%
\bibitem [{\citenamefont {Momma}\ and\ \citenamefont {Izumi}(2011)}]{VESTA}%
  \BibitemOpen
  \bibfield  {author} {\bibinfo {author} {\bibfnamefont {K.}~\bibnamefont
  {Momma}}\ and\ \bibinfo {author} {\bibfnamefont {F.}~\bibnamefont {Izumi}},\
  }\bibfield  {title} {\bibinfo {title} {{\it VESTA 3} for three-dimensional
  visualization of crystal, volumetric and morphology data},\ }\href
  {https://doi.org/10.1107/S0021889811038970} {\bibfield  {journal} {\bibinfo
  {journal} {J. Appl. Crystr.}\ }\textbf {\bibinfo {volume} {44}},\ \bibinfo
  {pages} {1272} (\bibinfo {year} {2011})}\BibitemShut {NoStop}%
\bibitem [{\citenamefont {Homes}\ \emph {et~al.}(1993)\citenamefont {Homes},
  \citenamefont {Reedyk}, \citenamefont {Crandles},\ and\ \citenamefont
  {Timusk}}]{Homes1993}%
  \BibitemOpen
  \bibfield  {author} {\bibinfo {author} {\bibfnamefont {C.~C.}\ \bibnamefont
  {Homes}}, \bibinfo {author} {\bibfnamefont {M.}~\bibnamefont {Reedyk}},
  \bibinfo {author} {\bibfnamefont {D.~A.}\ \bibnamefont {Crandles}},\ and\
  \bibinfo {author} {\bibfnamefont {T.}~\bibnamefont {Timusk}},\ }\bibfield
  {title} {\bibinfo {title} {Technique for measuring the reflectance of
  irregular, submillimeter-sized samples},\ }\href
  {https://doi.org/10.1364/AO.32.002976} {\bibfield  {journal} {\bibinfo
  {journal} {Appl. Opt.}\ }\textbf {\bibinfo {volume} {32}},\ \bibinfo {pages}
  {2976} (\bibinfo {year} {1993})}\BibitemShut {NoStop}%
\bibitem [{\citenamefont {Homes}\ \emph {et~al.}(2000)\citenamefont {Homes},
  \citenamefont {McConnell}, \citenamefont {Clayman}, \citenamefont {Bonn},
  \citenamefont {Liang}, \citenamefont {Hardy}, \citenamefont {Inoue},
  \citenamefont {Negishi}, \citenamefont {Fournier},\ and\ \citenamefont
  {Greene}}]{Homes2000}%
  \BibitemOpen
  \bibfield  {author} {\bibinfo {author} {\bibfnamefont {C.~C.}\ \bibnamefont
  {Homes}}, \bibinfo {author} {\bibfnamefont {A.~W.}\ \bibnamefont
  {McConnell}}, \bibinfo {author} {\bibfnamefont {B.~P.}\ \bibnamefont
  {Clayman}}, \bibinfo {author} {\bibfnamefont {D.~A.}\ \bibnamefont {Bonn}},
  \bibinfo {author} {\bibfnamefont {R.}~\bibnamefont {Liang}}, \bibinfo
  {author} {\bibfnamefont {W.~N.}\ \bibnamefont {Hardy}}, \bibinfo {author}
  {\bibfnamefont {M.}~\bibnamefont {Inoue}}, \bibinfo {author} {\bibfnamefont
  {H.}~\bibnamefont {Negishi}}, \bibinfo {author} {\bibfnamefont
  {P.}~\bibnamefont {Fournier}},\ and\ \bibinfo {author} {\bibfnamefont
  {R.~L.}\ \bibnamefont {Greene}},\ }\bibfield  {title} {\bibinfo {title}
  {{Phonon Screening in High-Temperature Superconductors}},\ }\href
  {https://doi.org/10.1103/PhysRevLett.84.5391} {\bibfield  {journal} {\bibinfo
   {journal} {Phys. Rev. Lett.}\ }\textbf {\bibinfo {volume} {84}},\ \bibinfo
  {pages} {5391} (\bibinfo {year} {2000})}\BibitemShut {NoStop}%
\bibitem [{Sup()}]{SupMat}%
  \BibitemOpen
  \href@noop {} {}\bibinfo {note} {See Supplemental Material at [URL will be
  inserted by publisher] for a discussion of how a weak electronic background
  can fundamentally alter the vibrational line shape in the reflectivity.
  Examples of terracing in EuCd$_2$P$_2$ are shown; the experimental
  high-resolution (0.2~cm$^{-1}$) in-plane [(001)] reflectivty, as well as the
  reflectivity from an \emph{a-c} [(101)] face is shown. The fitted values of
  the in-plane $E_u$ and \emph{c}-axis $A_{2u}$ modes, as well as estimates for
  the positions of the LO modes, are presented; estimates of $\epsilon_\infty$
  are validated by examining the positions of the \emph{a-b} plane TO and LO
  modes. The analysis concludes that terracing leads to artifacts in the
  reflectivity due to the simple mixing of the \emph{a-b} plane and \emph{c}
  axis reflectivity.}\BibitemShut {Stop}%
\bibitem [{\citenamefont {Dressel}\ and\ \citenamefont
  {Gr{\"u}ner}(2001)}]{Dressel-Book}%
  \BibitemOpen
  \bibfield  {author} {\bibinfo {author} {\bibfnamefont {M.}~\bibnamefont
  {Dressel}}\ and\ \bibinfo {author} {\bibfnamefont {G.}~\bibnamefont
  {Gr{\"u}ner}},\ }\href@noop {} {\emph {\bibinfo {title} {Electrodynamics of
  Solids}}}\ (\bibinfo  {publisher} {Cambridge University Press},\ \bibinfo
  {address} {Cambridge},\ \bibinfo {year} {2001})\BibitemShut {NoStop}%
\bibitem [{\citenamefont {Wooten}(1972)}]{Wooten}%
  \BibitemOpen
  \bibfield  {author} {\bibinfo {author} {\bibfnamefont {F.}~\bibnamefont
  {Wooten}},\ }\href@noop {} {\emph {\bibinfo {title} {Optical Properties of
  Solids}}}\ (\bibinfo  {publisher} {Academic Press},\ \bibinfo {address} {New
  York},\ \bibinfo {year} {1972})\ pp.\ \bibinfo {pages} {244--250}\BibitemShut
  {NoStop}%
\bibitem [{\citenamefont {Fano}(1961)}]{Fano1961}%
  \BibitemOpen
  \bibfield  {author} {\bibinfo {author} {\bibfnamefont {U.}~\bibnamefont
  {Fano}},\ }\bibfield  {title} {\bibinfo {title} {Effects of configuration
  interaction on intensities and phase shifts},\ }\href
  {https://doi.org/10.1103/PhysRev.124.1866} {\bibfield  {journal} {\bibinfo
  {journal} {Phys. Rev.}\ }\textbf {\bibinfo {volume} {124}},\ \bibinfo {pages}
  {1866} (\bibinfo {year} {1961})}\BibitemShut {NoStop}%
\bibitem [{\citenamefont {Damascelli}(1996)}]{Damascelli1996}%
  \BibitemOpen
  \bibfield  {author} {\bibinfo {author} {\bibfnamefont {A.}~\bibnamefont
  {Damascelli}},\ }\emph {\bibinfo {title} {Optical Spectroscopy of Quantum
  Spin Systems}},\ \href@noop {} {Ph.D. thesis},\ \bibinfo  {school}
  {University of Groningen} (\bibinfo {year} {1996}),\ \bibinfo {note}
  {p.~21}\BibitemShut {NoStop}%
\bibitem [{\citenamefont {Wang}\ \emph {et~al.}(2017)\citenamefont {Wang},
  \citenamefont {Rademaker}, \citenamefont {Dagotto},\ and\ \citenamefont
  {Johnston}}]{Wang2017}%
  \BibitemOpen
  \bibfield  {author} {\bibinfo {author} {\bibfnamefont {Y.}~\bibnamefont
  {Wang}}, \bibinfo {author} {\bibfnamefont {L.}~\bibnamefont {Rademaker}},
  \bibinfo {author} {\bibfnamefont {E.}~\bibnamefont {Dagotto}},\ and\ \bibinfo
  {author} {\bibfnamefont {S.}~\bibnamefont {Johnston}},\ }\bibfield  {title}
  {\bibinfo {title} {{Phonon linewidth due to electron-phonon interactions with
  strong forward scattering in FeSe thin films on oxide substrates}},\ }\href
  {https://doi.org/10.1103/PhysRevB.96.054515} {\bibfield  {journal} {\bibinfo
  {journal} {Phys. Rev. B}\ }\textbf {\bibinfo {volume} {96}},\ \bibinfo
  {pages} {054515} (\bibinfo {year} {2017})}\BibitemShut {NoStop}%
\bibitem [{\citenamefont {Klemens}(1966)}]{Klemens1966}%
  \BibitemOpen
  \bibfield  {author} {\bibinfo {author} {\bibfnamefont {P.~G.}\ \bibnamefont
  {Klemens}},\ }\bibfield  {title} {\bibinfo {title} {Anharmonic decay of
  optical phonons},\ }\href {https://doi.org/10.1103/PhysRev.148.845}
  {\bibfield  {journal} {\bibinfo  {journal} {Phys. Rev.}\ }\textbf {\bibinfo
  {volume} {148}},\ \bibinfo {pages} {845} (\bibinfo {year}
  {1966})}\BibitemShut {NoStop}%
\bibitem [{\citenamefont {Men\'endez}\ and\ \citenamefont
  {Cardona}(1984)}]{Menendez1984}%
  \BibitemOpen
  \bibfield  {author} {\bibinfo {author} {\bibfnamefont {J.}~\bibnamefont
  {Men\'endez}}\ and\ \bibinfo {author} {\bibfnamefont {M.}~\bibnamefont
  {Cardona}},\ }\bibfield  {title} {\bibinfo {title} {{Temperature dependence
  of the first-order Raman scattering by phonons in Si, Ge, and
  $\ensuremath{\alpha}-\mathrm{S}\mathrm{n}$: Anharmonic effects}},\ }\href
  {https://doi.org/10.1103/PhysRevB.29.2051} {\bibfield  {journal} {\bibinfo
  {journal} {Phys. Rev. B}\ }\textbf {\bibinfo {volume} {29}},\ \bibinfo
  {pages} {2051} (\bibinfo {year} {1984})}\BibitemShut {NoStop}%
\bibitem [{\citenamefont {Homes}\ \emph {et~al.}(2016)\citenamefont {Homes},
  \citenamefont {Dai}, \citenamefont {Schneeloch}, \citenamefont {Zhong},\ and\
  \citenamefont {Gu}}]{Homes2016}%
  \BibitemOpen
  \bibfield  {author} {\bibinfo {author} {\bibfnamefont {C.~C.}\ \bibnamefont
  {Homes}}, \bibinfo {author} {\bibfnamefont {Y.~M.}\ \bibnamefont {Dai}},
  \bibinfo {author} {\bibfnamefont {J.}~\bibnamefont {Schneeloch}}, \bibinfo
  {author} {\bibfnamefont {R.~D.}\ \bibnamefont {Zhong}},\ and\ \bibinfo
  {author} {\bibfnamefont {G.~D.}\ \bibnamefont {Gu}},\ }\bibfield  {title}
  {\bibinfo {title} {Phonon anomalies in some iron telluride materials},\
  }\href {https://doi.org/10.1103/PhysRevB.93.125135} {\bibfield  {journal}
  {\bibinfo  {journal} {Phys. Rev. B}\ }\textbf {\bibinfo {volume} {93}},\
  \bibinfo {pages} {125135} (\bibinfo {year} {2016})}\BibitemShut {NoStop}%
\bibitem [{\citenamefont {Berreman}(1970)}]{Berreman1970}%
  \BibitemOpen
  \bibfield  {author} {\bibinfo {author} {\bibfnamefont {D.~W.}\ \bibnamefont
  {Berreman}},\ }\bibfield  {title} {\bibinfo {title} {{Resonant Reflectance
  Anomalies: Effect of Shapes of Surface Irregularities}},\ }\href
  {https://doi.org/10.1103/PhysRevB.1.381} {\bibfield  {journal} {\bibinfo
  {journal} {Phys. Rev. B}\ }\textbf {\bibinfo {volume} {1}},\ \bibinfo {pages}
  {381} (\bibinfo {year} {1970})}\BibitemShut {NoStop}%
\bibitem [{\citenamefont {Reedyk}\ and\ \citenamefont
  {Timusk}(1992)}]{Reedyk1992}%
  \BibitemOpen
  \bibfield  {author} {\bibinfo {author} {\bibfnamefont {M.}~\bibnamefont
  {Reedyk}}\ and\ \bibinfo {author} {\bibfnamefont {T.}~\bibnamefont
  {Timusk}},\ }\bibfield  {title} {\bibinfo {title} {{Evidence for
  \emph{a-b}-plane coupling to longitudinal \emph{c}-axis phonons in high-$T_c$
  superconductors}},\ }\href {https://doi.org/10.1103/PhysRevLett.69.2705}
  {\bibfield  {journal} {\bibinfo  {journal} {Phys. Rev. Lett.}\ }\textbf
  {\bibinfo {volume} {69}},\ \bibinfo {pages} {2705} (\bibinfo {year}
  {1992})}\BibitemShut {NoStop}%
\bibitem [{\citenamefont {Homes}\ \emph {et~al.}(2007)\citenamefont {Homes},
  \citenamefont {Tranquada},\ and\ \citenamefont {Buttrey}}]{Homes2007}%
  \BibitemOpen
  \bibfield  {author} {\bibinfo {author} {\bibfnamefont {C.~C.}\ \bibnamefont
  {Homes}}, \bibinfo {author} {\bibfnamefont {J.~M.}\ \bibnamefont
  {Tranquada}},\ and\ \bibinfo {author} {\bibfnamefont {D.~J.}\ \bibnamefont
  {Buttrey}},\ }\bibfield  {title} {\bibinfo {title} {{Stripe order and
  vibrational properties of La$_{2}$NiO$_{4+\delta}$ for $\delta=2/15$:
  Measurements and \emph{ab initio} calculations}},\ }\href
  {https://doi.org/10.1103/PhysRevB.75.045128} {\bibfield  {journal} {\bibinfo
  {journal} {Phys. Rev. B}\ }\textbf {\bibinfo {volume} {75}},\ \bibinfo
  {pages} {045128} (\bibinfo {year} {2007})}\BibitemShut {NoStop}%
\bibitem [{\citenamefont {Emin}(1993)}]{Emin1993}%
  \BibitemOpen
  \bibfield  {author} {\bibinfo {author} {\bibfnamefont {D.}~\bibnamefont
  {Emin}},\ }\bibfield  {title} {\bibinfo {title} {Optical properties of large
  and small polarons and bipolarons},\ }\href
  {https://doi.org/10.1103/PhysRevB.48.13691} {\bibfield  {journal} {\bibinfo
  {journal} {Phys. Rev. B}\ }\textbf {\bibinfo {volume} {48}},\ \bibinfo
  {pages} {13691} (\bibinfo {year} {1993})}\BibitemShut {NoStop}%
\bibitem [{\citenamefont {Alexandrov}\ and\ \citenamefont
  {Mott}(1995)}]{Alexandrov-Book}%
  \BibitemOpen
  \bibfield  {author} {\bibinfo {author} {\bibfnamefont {A.~S.}\ \bibnamefont
  {Alexandrov}}\ and\ \bibinfo {author} {\bibfnamefont {N.}~\bibnamefont
  {Mott}},\ }\href@noop {} {\emph {\bibinfo {title} {Polarons \& Bipolarons}}}\
  (\bibinfo  {publisher} {World Scientific},\ \bibinfo {address} {Cambridge},\
  \bibinfo {year} {1995})\BibitemShut {NoStop}%
\bibitem [{\citenamefont {Millis}\ \emph {et~al.}(1996)\citenamefont {Millis},
  \citenamefont {Mueller},\ and\ \citenamefont {Shraiman}}]{Millis1996}%
  \BibitemOpen
  \bibfield  {author} {\bibinfo {author} {\bibfnamefont {A.~J.}\ \bibnamefont
  {Millis}}, \bibinfo {author} {\bibfnamefont {R.}~\bibnamefont {Mueller}},\
  and\ \bibinfo {author} {\bibfnamefont {B.~I.}\ \bibnamefont {Shraiman}},\
  }\bibfield  {title} {\bibinfo {title} {{Fermi-liquid-to-polaron crossover.
  II. Double exchange and the physics of colossal magnetoresistance}},\ }\href
  {https://doi.org/10.1103/PhysRevB.54.5405} {\bibfield  {journal} {\bibinfo
  {journal} {Phys. Rev. B}\ }\textbf {\bibinfo {volume} {54}},\ \bibinfo
  {pages} {5405} (\bibinfo {year} {1996})}\BibitemShut {NoStop}%
\bibitem [{\citenamefont {R\"oder}\ \emph {et~al.}(1996)\citenamefont
  {R\"oder}, \citenamefont {Zang},\ and\ \citenamefont {Bishop}}]{Roder1996}%
  \BibitemOpen
  \bibfield  {author} {\bibinfo {author} {\bibfnamefont {H.}~\bibnamefont
  {R\"oder}}, \bibinfo {author} {\bibfnamefont {J.}~\bibnamefont {Zang}},\ and\
  \bibinfo {author} {\bibfnamefont {A.~R.}\ \bibnamefont {Bishop}},\ }\bibfield
   {title} {\bibinfo {title} {{Lattice Effects in the
  Colossal-Magnetoresistance Manganites}},\ }\href
  {https://doi.org/10.1103/PhysRevLett.76.1356} {\bibfield  {journal} {\bibinfo
   {journal} {Phys. Rev. Lett.}\ }\textbf {\bibinfo {volume} {76}},\ \bibinfo
  {pages} {1356} (\bibinfo {year} {1996})}\BibitemShut {NoStop}%
\bibitem [{\citenamefont {Lee}\ and\ \citenamefont {Min}(1997)}]{Lee1997}%
  \BibitemOpen
  \bibfield  {author} {\bibinfo {author} {\bibfnamefont {J.~D.}\ \bibnamefont
  {Lee}}\ and\ \bibinfo {author} {\bibfnamefont {B.~I.}\ \bibnamefont {Min}},\
  }\bibfield  {title} {\bibinfo {title} {Polaron transport and lattice dynamics
  in colossal-magnetoresistance manganites},\ }\href
  {https://doi.org/10.1103/PhysRevB.55.12454} {\bibfield  {journal} {\bibinfo
  {journal} {Phys. Rev. B}\ }\textbf {\bibinfo {volume} {55}},\ \bibinfo
  {pages} {12454} (\bibinfo {year} {1997})}\BibitemShut {NoStop}%
\bibitem [{\citenamefont {Salamon}\ and\ \citenamefont
  {Jaime}(2001)}]{Salamon2001}%
  \BibitemOpen
  \bibfield  {author} {\bibinfo {author} {\bibfnamefont {M.~B.}\ \bibnamefont
  {Salamon}}\ and\ \bibinfo {author} {\bibfnamefont {M.}~\bibnamefont
  {Jaime}},\ }\bibfield  {title} {\bibinfo {title} {{The physics of manganites:
  Structure and transport}},\ }\href
  {https://doi.org/10.1103/RevModPhys.73.583} {\bibfield  {journal} {\bibinfo
  {journal} {Rev. Mod. Phys.}\ }\textbf {\bibinfo {volume} {73}},\ \bibinfo
  {pages} {583} (\bibinfo {year} {2001})}\BibitemShut {NoStop}%
\bibitem [{\citenamefont {Lanzara}\ \emph {et~al.}(1998)\citenamefont
  {Lanzara}, \citenamefont {Saini}, \citenamefont {Brunelli}, \citenamefont
  {Natali}, \citenamefont {Bianconi}, \citenamefont {Radaelli},\ and\
  \citenamefont {Cheong}}]{Lanzara1998}%
  \BibitemOpen
  \bibfield  {author} {\bibinfo {author} {\bibfnamefont {A.}~\bibnamefont
  {Lanzara}}, \bibinfo {author} {\bibfnamefont {N.~L.}\ \bibnamefont {Saini}},
  \bibinfo {author} {\bibfnamefont {M.}~\bibnamefont {Brunelli}}, \bibinfo
  {author} {\bibfnamefont {F.}~\bibnamefont {Natali}}, \bibinfo {author}
  {\bibfnamefont {A.}~\bibnamefont {Bianconi}}, \bibinfo {author}
  {\bibfnamefont {P.~G.}\ \bibnamefont {Radaelli}},\ and\ \bibinfo {author}
  {\bibfnamefont {S.-W.}\ \bibnamefont {Cheong}},\ }\bibfield  {title}
  {\bibinfo {title} {{Crossover from Large to Small Polarons across the
  Metal-Insulator Transition in Manganites}},\ }\href
  {https://doi.org/10.1103/PhysRevLett.81.878} {\bibfield  {journal} {\bibinfo
  {journal} {Phys. Rev. Lett.}\ }\textbf {\bibinfo {volume} {81}},\ \bibinfo
  {pages} {878} (\bibinfo {year} {1998})}\BibitemShut {NoStop}%
\bibitem [{\citenamefont {Hartinger}\ \emph {et~al.}(2004)\citenamefont
  {Hartinger}, \citenamefont {Mayr}, \citenamefont {Deisenhofer}, \citenamefont
  {Loidl},\ and\ \citenamefont {Kopp}}]{Hartinger2004}%
  \BibitemOpen
  \bibfield  {author} {\bibinfo {author} {\bibfnamefont {C.}~\bibnamefont
  {Hartinger}}, \bibinfo {author} {\bibfnamefont {F.}~\bibnamefont {Mayr}},
  \bibinfo {author} {\bibfnamefont {J.}~\bibnamefont {Deisenhofer}}, \bibinfo
  {author} {\bibfnamefont {A.}~\bibnamefont {Loidl}},\ and\ \bibinfo {author}
  {\bibfnamefont {T.}~\bibnamefont {Kopp}},\ }\bibfield  {title} {\bibinfo
  {title} {{Large and small polaron excitations in
  La$_{2/3}$(Sr/Ca)$_{1/3}$MnO$_3$ films}},\ }\href
  {https://doi.org/10.1103/PhysRevB.69.100403} {\bibfield  {journal} {\bibinfo
  {journal} {Phys. Rev. B}\ }\textbf {\bibinfo {volume} {69}},\ \bibinfo
  {pages} {100403(R)} (\bibinfo {year} {2004})}\BibitemShut {NoStop}%
\bibitem [{\citenamefont {Zhang}\ \emph {et~al.}(2010)\citenamefont {Zhang},
  \citenamefont {Chen}, \citenamefont {He}, \citenamefont {Yang}, \citenamefont
  {Xie}, \citenamefont {Xie}, \citenamefont {Chen}, \citenamefont {Fang},
  \citenamefont {Arita}, \citenamefont {Shimada}, \citenamefont {Namatame},
  \citenamefont {Taniguchi}, \citenamefont {Hu},\ and\ \citenamefont
  {Feng}}]{Zhang2010}%
  \BibitemOpen
  \bibfield  {author} {\bibinfo {author} {\bibfnamefont {Y.}~\bibnamefont
  {Zhang}}, \bibinfo {author} {\bibfnamefont {F.}~\bibnamefont {Chen}},
  \bibinfo {author} {\bibfnamefont {C.}~\bibnamefont {He}}, \bibinfo {author}
  {\bibfnamefont {L.~X.}\ \bibnamefont {Yang}}, \bibinfo {author}
  {\bibfnamefont {B.~P.}\ \bibnamefont {Xie}}, \bibinfo {author} {\bibfnamefont
  {Y.~L.}\ \bibnamefont {Xie}}, \bibinfo {author} {\bibfnamefont {X.~H.}\
  \bibnamefont {Chen}}, \bibinfo {author} {\bibfnamefont {M.}~\bibnamefont
  {Fang}}, \bibinfo {author} {\bibfnamefont {M.}~\bibnamefont {Arita}},
  \bibinfo {author} {\bibfnamefont {K.}~\bibnamefont {Shimada}}, \bibinfo
  {author} {\bibfnamefont {H.}~\bibnamefont {Namatame}}, \bibinfo {author}
  {\bibfnamefont {M.}~\bibnamefont {Taniguchi}}, \bibinfo {author}
  {\bibfnamefont {J.~P.}\ \bibnamefont {Hu}},\ and\ \bibinfo {author}
  {\bibfnamefont {D.~L.}\ \bibnamefont {Feng}},\ }\bibfield  {title} {\bibinfo
  {title} {{Strong correlations and spin-density-wave phase induced by a
  massive spectral weight redistribution in $\alpha-$Fe$_{1.06}$Te}},\ }\href
  {https://doi.org/10.1103/PhysRevB.82.165113} {\bibfield  {journal} {\bibinfo
  {journal} {Phys. Rev. B}\ }\textbf {\bibinfo {volume} {82}},\ \bibinfo
  {pages} {165113} (\bibinfo {year} {2010})}\BibitemShut {NoStop}%
\end{thebibliography}
%
%

%
%

\clearpage
\newpage

\newpage
\vspace*{-2.1cm}
\hspace*{-2.5cm}
{
  \centering
  \includegraphics[width=1.2\textwidth,page=1]{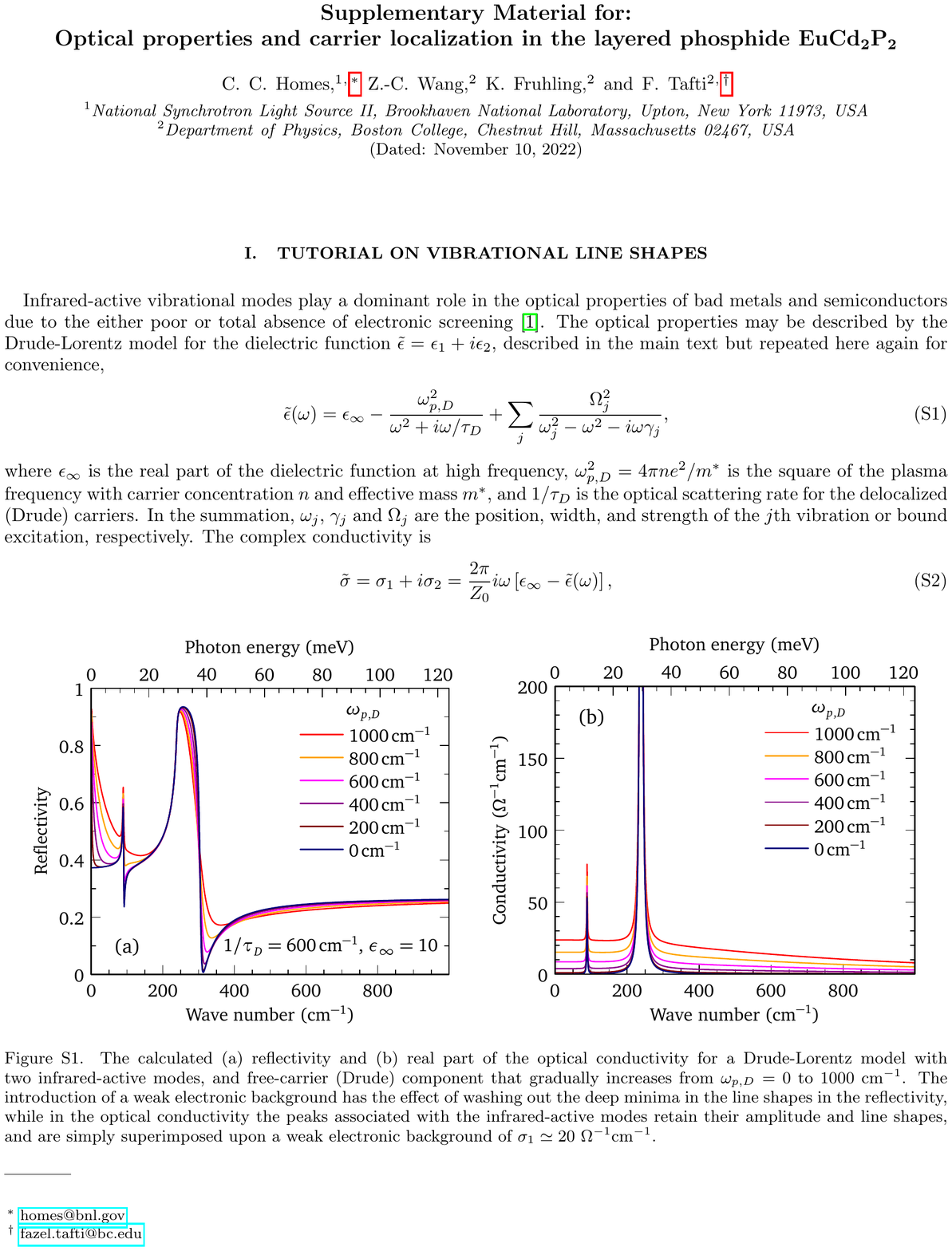} \\
  \ \\
}

\newpage
\vspace*{-2.1cm}
\hspace*{-2.5cm}
{
  \centering
  \includegraphics[width=1.2\textwidth,page=2]{supplemental.pdf} \\
  \ \\
}

\newpage
\vspace*{-2.1cm}
\hspace*{-2.5cm}
{
  \centering
  \includegraphics[width=1.2\textwidth,page=3]{supplemental.pdf} \\
  \ \\
}

\newpage
\vspace*{-2.1cm}
\hspace*{-2.5cm}
{
  \centering
  \includegraphics[width=1.2\textwidth,page=4]{supplemental.pdf} \\
  \ \\
}

\newpage
\vspace*{-2.1cm}
\hspace*{-2.5cm}
{
  \centering
  \includegraphics[width=1.2\textwidth,page=5]{supplemental.pdf} \\
  \ \\
}

\end{document}